\documentclass[twocolumn]{aastex62}

\usepackage{url}
\newcommand{\sourcefile}{\path}
\usepackage{amsmath}
\usepackage{rotating}
\graphicspath{{./}{figures/}}

\shorttitle{Exploring Trans-Neptunian Space with TESS}
\shortauthors{Rice \& Laughlin}

\begin{document}

\title{Exploring Trans-Neptunian Space with TESS: A Targeted Shift-Stacking Search for Planet Nine and Distant TNOs in the Galactic Plane}

\author[0000-0002-7670-670X]{Malena Rice}
\affiliation{Department of Astronomy, Yale University, New Haven, CT 06511, USA}
\affiliation{NSF Graduate Research Fellow}

\author{Gregory Laughlin}
\affiliation{Department of Astronomy, Yale University, New Haven, CT 06511, USA}

\correspondingauthor{Malena Rice}
\email{malena.rice@yale.edu}

\begin{abstract}
We present results from a new pipeline custom-designed to search for faint, undiscovered solar system bodies using full-frame image data from the NASA Transiting Exoplanet Survey Satellite (TESS) mission. This pipeline removes the baseline flux of each pixel before aligning and co-adding frames along plausible orbital paths of interest. We first demonstrate the performance of the pipeline by recovering the signals of three trans-Neptunian objects -- 90377 Sedna ($V=20.64$), 2015 BP519 ($V=21.81$), and 2007 TG422 ($V=22.32$) -- both through shift-stacking along their known sky-projected paths and through a blind recovery. We then apply this blind search procedure in a proof-of-concept survey of TESS Sectors 18 and 19, which extend through a portion of the galactic plane in the Northern Hemisphere. We search for dim objects at geocentric distances $d=70-800$ au in a targeted search for Planet Nine and any previously unknown detached Kuiper belt objects that may shed light on the Planet Nine hypothesis. With no input orbital information, our present pipeline can reliably recover the signals of distant solar system bodies in the galactic plane with $V<21$ and current distances $d\lesssim 150$ au, and we elaborate on paths forward to push these limits in future optimizations. The methods described in this paper will serve as a foundation for an all-sky shift-stacking survey of the distant solar system with TESS.
\end{abstract}


\section{Introduction} 
\label{section:intro}

The outer reaches of the solar system, at distances $d\geq70$ au, remain largely unexplored. Individual objects with orbits exterior to Neptune's -- known as trans-Neptunian objects, or TNOs -- are challenging to detect; owing to the steep, $F \propto 1/r^4$ diminution of reflected flux, only the brightest such objects have been discovered. Indeed, at the time of writing, fewer than 100 detached Kuiper belt objects (KBOs), with perihelia $q \gtrsim 40$ au and no direct interactions with the known solar system planets, have been discovered.

These objects, however, are of exceptional interest due to the unique window that they provide into the dynamical evolution of the outer solar system. The observed apsidal and nodal alignment of detached KBOs, combined with the unexpectedly high inclinations and eccentricities of several outer solar system small body populations, has led to the proposition that a ninth, super-Earth sized planet, commonly known as ``Planet Nine", may reside in the distant solar system \citep{trujillo2014, batygin2016, malhotra2016corralling}. Dynamical simulations reveal that, in order to account for this observed alignment among KBOs, Planet Nine would likely possess a large semimajor axis ($a=400-800$ au), a high inclination ($i=15-25\degr$), and a substantial eccentricity \citep[$e=0.2-0.5$; ][]{batygin2019planet}. The best-fit orbital solution for Planet Nine found by \citet{batygin2019planet}, with $a = 500$ au and aphelion distance $Q=625$ au, corresponds to magnitude $V=19.0-22.2$, where the exact value is determined by the planet's location along the orbit and its inferred albedo.

Several recent and ongoing searches have placed observational constraints on the properties of Planet Nine. \citet{meisner2018} applied data from the Wide-Field Infrared Survey Explorer \citep[WISE;][]{wright2010} to search for the proposed planet at high galactic latitudes, ruling out a bright planet ($W1 < 16.7$, where the $W1$ bandpass is centered at 3.4 $\mu$m) at 90\% completeness over $3\pi$ radians on the sky. The Backyard Worlds: Planet 9 citizen science project described in \citet{kuchner2017first} has also used the WISE dataset in a more extended Planet Nine search, employing $W1$ along with an additional wavelength band ($W2$, centered at 4.6 $\mu$m) and resulting in the discovery of the brown dwarf WISEA J110125.95+540052.8. While it is not specifically designed to search for Planet-Nine-like signals, the Dark Energy Survey \citep[DES;][]{dark2005dark} covers 5000 square degrees in the southern sky and is sensitive to dim Planet Nine-like signals at optical and near-infared wavelengths. From the survey's first four years on-sky, the DES collaboration reported the discovery of hundreds of TNOs, including some relevant to the Planet Nine hypothesis \citep{bernardinelli2020trans, becker2018discovery}. Indirect, gravitational searches have also been pursued \citep{fienga2016, holman2016observationalc, holman2016observationalp}; however, these searches require a long time baseline of precise positional measurements across many gravitational probes to distinguish the effects of Planet Nine from those induced by the Kuiper belt \citep{rice2019case}.

Planet Nine remains elusive. If it exists, the most promising places to look might now be the regions of the sky with the highest stellar density, where a slowly moving point source is particularly difficult to pinpoint and isolate. A bright and compelling signal lost in the noise of the galactic plane would not be unprecedented; for example, the binary brown dwarf system Luhman 16, only 2 pc from the Sun, remained undiscovered until 2013 as a consequence of its proximity to the galactic plane \citep{luhman2013discovery}.

We present results from a systematic search for objects in the outer solar system ($a = 70-800$ au) using a custom shift-stacking pipeline designed for use with full-frame images (FFIs) from the Transiting Exoplanet Survey Satellite \citep[TESS;][]{ricker2015tess}. The basic idea of the underlying shift-stacking technique -- also called ``digital tracking", ``de-orbiting", or ``pencil-beam surveys" in the literature -- has been implemented in several preceding works to search for new solar system satellites \citep{holman2004discovery, kavelaars2004discovery, burkhart2016deep} and TNOs \citep{gladman1998pencil, gladman2001structure, bernstein2004size}, and a version of it was recently proposed for application to TESS by \citet{holman2019tess}. Our implementation includes a number of refinements that are specific to finding particularly distant solar system objects in the TESS dataset.

For the purposes of this study, we focus on Sectors 18 and 19, which lie directly along the galactic plane. Our focus on this region is motivated by two factors. First, based on the most recent parameter updates provided by \citet{batygin2019planet}, the most likely remaining parameter space for Planet Nine -- and, specifically, the parameter space that is most poorly constrained by other surveys such as Pan-STARRS \citep{kaiser2002pan} -- lies in the vicinity of the galactic plane (see Figure 25 of \citet{batygin2019planet}). If Planet Nine exists in the galactic plane, this would help to explain why it has not yet been discovered, since stellar contamination severely limits optical searches in this region of the sky. Thus, by focusing on Sectors 18 and 19, which encompass much of the galactic plane in the northern sky, we complete a targeted search aimed at the region in which Planet Nine is most likely to lie. 

Second, a survey of Sectors 18 and 19 allows us to quantify the performance of our pipeline in a noisy region of the sky that has previously been difficult to study with single-frame exposures. We demonstrate that shift-stacking is a promising method to search for dim outer solar system objects using all-sky surveys, strengthened by its ability to recover sources that would otherwise be lost in the noise due to stellar crowding in single-frame exposures. While previous studies have employed the TESS dataset for solar system science applications \citep{pal2018tess, mcneill2019asteroid, pal2020solar}, we present results from the first survey designed to detect undiscovered solar system objects in the TESS dataset.

\section{Data Overview}
\label{section:data_overview}
The TESS spacecraft includes 4 cameras each comprised of 4 CCDs with 2048 x 2048 pixels per CCD. Each pixel spans $21\arcsec\times21\arcsec$ for a combined, total field of view $24\degr\times96\degr$, extending from the ecliptic pole towards the ecliptic plane at each spacecraft pointing. The survey's observing strategy covers most of the sky away from the ecliptic plane, extending down to ecliptic latitude $b\gtrsim 6\degr$ and spending the most time observing the ecliptic poles (the ``continuous viewing zone" that retains coverage as the pointings change). TESS is thus ideally suited to search for high-inclination solar system objects. Due to the TESS camera's point spread function (PSF), 50\% of the flux from an object falls within 1 pix$^2$ of the object's peak flux location while 90\% falls within 4 pix$^2$.\footnote{\url{https://heasarc.gsfc.nasa.gov/docs/tess/the-tess-space-telescope.html}}

TESS observations are organized into sectors, each of which corresponds to a single spacecraft pointing. Each hemisphere is spanned by thirteen observing sectors, and TESS spends 27 consecutive days collecting data for each sector. These observations include 30-minute cadence full-frame images over the full field of view, as well as 2-minute cadence ``postage stamp" images of individual bright stars. 

\begin{figure*}
    \centering
    \includegraphics[width=0.98\linewidth]{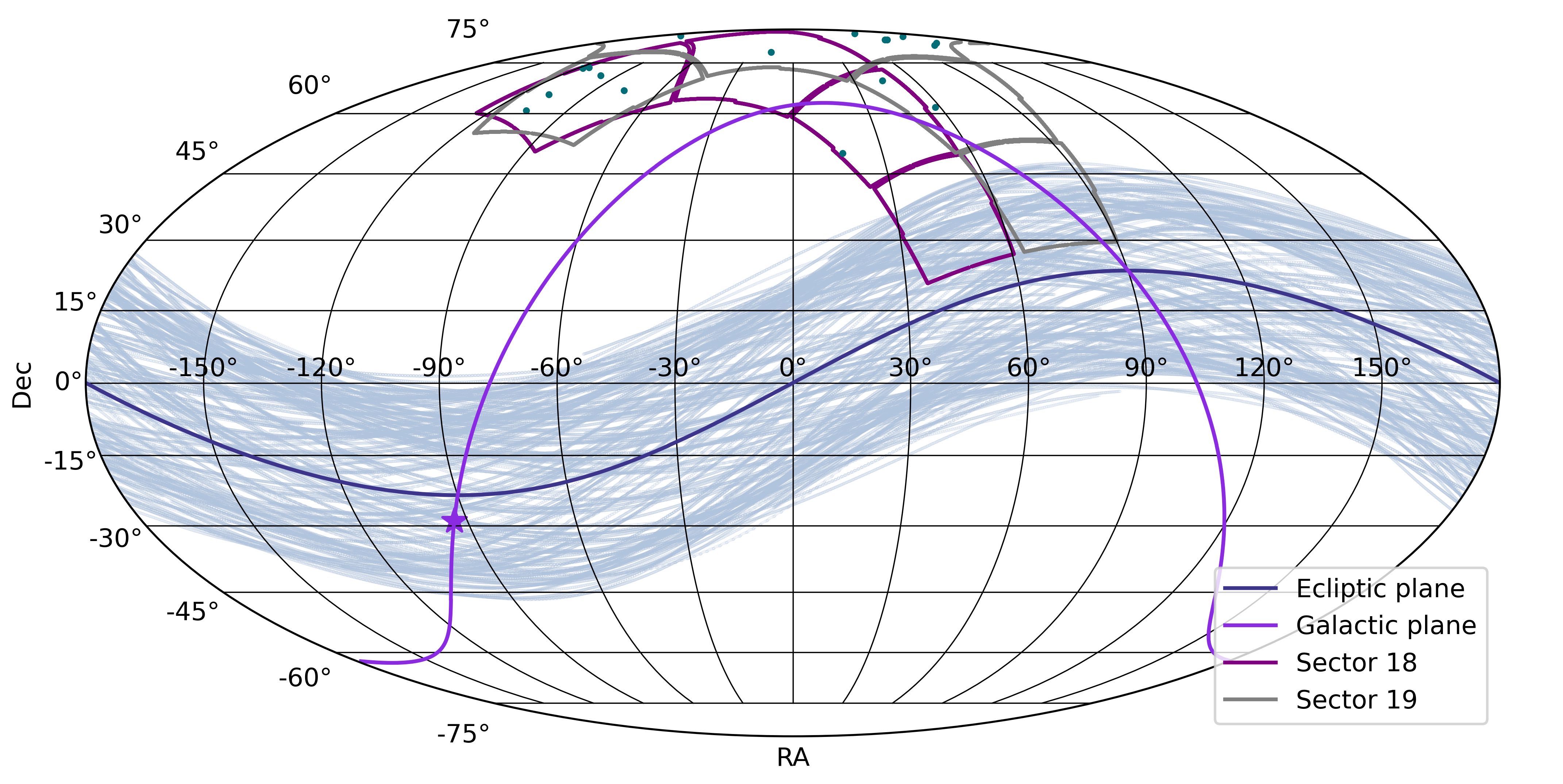}
    \caption{Mollweide projection of the TESS field of view in celestial coordinates, with the ecliptic plane shown in navy and the galactic plane in purple. A purple star denotes the location of the galactic center. The regions of the sky observed in Sectors 18 and 19 are delineated in red and gray, respectively. We sample and plot 150 random Planet Nine orbits in light blue to show the overlap between the possible Planet Nine orbits and the region observed by TESS. We target the region of space in which the galactic plane intersects with possible Planet Nine orbits. The candidate outer solar system objects presented in this study are included in green for reference.}
    \label{fig:TESS_FOV}
\end{figure*}

We take advantage of TESS's extensive sky coverage by searching for dim, slow-moving objects in the calibrated Sector 18 and Sector 19 FFIs. The locations of these sectors in the sky relative to the ecliptic plane, the galactic plane, and the range of expected Planet Nine orbits are displayed in Figure \ref{fig:TESS_FOV}. The sampled Planet Nine orbital elements were randomly selected from the ranges $400$ au $<a<800$ au, $15\degr \leq i \leq 25\degr$, and $0.2\leq e \leq 0.5$ with randomly oriented orbital angles. We then used the \texttt{PyEphem} software package to translate the selected orbital elements to projected sky locations and to plot those locations over a span of 10,000 years (the limiting time span covered by \texttt{PyEphem}).

To quantify the performance of our pipeline, we also recover previously known objects using calibrated FFIs from Sector 5. The difference in flux density based on proximity to the galactic plane is apparent: the average pixel's flux in Camera 1 of Sector 19, which lies along the galactic plane, is a factor of $\sim$1.5 higher than in Camera 1 of Sector 5, which is at a $\sim30\degr$ separation from the galactic plane. We download all frames, which have been corrected for instrument/detector artifacts, directly from the Mikulski Archive for Space Telescopes (MAST).\footnote{\url{http://archive.stsci.edu/tess/bulk_downloads/bulk_downloads_ffi-tp-lc-dv.html}}

\section{Methods}
\label{section:methods}
To search for TNOs in the TESS dataset, we developed a linear shift-stacking pipeline that checks a pre-defined set of potential orbit paths for objects of interest. This pipeline includes three primary components: (1) an algorithm that subtracts the flux baseline on a per-pixel basis, (2) a shift-stacking algorithm that co-adds all baseline-subtracted images along potential TNO paths in search of strong local maxima, and (3) an automated algorithm for extracting candidates. By summing many small deviations from the flux baseline over a large number of exposures, we recover dim objects far below the detection limit of an individual frame.

We divided each TESS CCD into $256\times256$ pixel cutout regions in a grid that overlaps itself by 128 pixels in either direction. With this partition, each region of the sky interior to the edge of the CCD is present in four different shift-stacking frames. This degeneracy improves the likelihood that any given frame will enclose the full path of an outer solar system object. It also decreases the likelihood that we will miss any given object during our vetting procedure.

We ran our pipeline on the Grace cluster at the Yale Center for Research Computing, using one compute node and one CPU per $256\times256$ pixel cutout region. The full pipeline described in this section takes 1-3 hours (wall time) to run with the polynomial baseline subtraction, while this time is increased to 7-8 hours with the PCA baseline subtraction. We processed each cutout frame independently; as a result, we were able to run a full sector at a time, processing all cutout regions in parallel.

\subsection{Baseline Subtraction Algorithms}
Our baseline subtraction procedure includes both a pre-processing pixel masking stage (Section \ref{section:prefit_pix_mask}) and two independent methods for removing the flux baseline, where both methods fit the baseline on a pixel-by-pixel basis. We use the polynomial baseline subtraction method described in Section \ref{section:poly_baseline_sub} to return our key results, and we use the Principal Component Analysis (PCA) baseline subtraction method described in Section \ref{section:pca_baseline_sub} as a consistency check. By probing parameter space with two independent data processing procedures, we ensure that only the most robust candidates remain in our final search results.

\subsubsection{Pre-Fit Pixel Masking}
\label{section:prefit_pix_mask}
We first use a predetermined mask, set individually for each sector during a preliminary testing phase, to remove a fraction of frames displaying large systematic signals from the full time series. These masks eliminate flux discontinuities in the pixel time series that generally occur at the beginning, middle, and/or end of each TESS sector. In Sectors 18 and 19, the removal of discontinuities at the center of each time series leaves two roughly equal-length $\sim6-7$ day light curves for each pixel, separated by a $\sim5-10$ day gap.

The two time-series masks -- one for Sector 18 and another for Sector 19 --  were selected using an iterative trial-and-error process with our polynomial baseline subtraction method (Section \ref{section:poly_baseline_sub}). For each sector, we obtain a mask that removes the fewest possible time frames while still providing a consistently well-performing fit to each of the two light curve segments. We examine the residuals of our polynomial fits in search of asymmetric systematics and alter our masks accordingly.

After this first pass at removing discontinuities, we also remove the 10\% of time series frames where the median flux gradient across all pixels has the largest magnitude. In this way, we discard systematics corresponding to rapid flux changes affecting the full cutout region, which introduce additional scatter and degrade the quality of our fits. This allows us to more accurately determine the baseline background flux present in all pixels.

We then determine the maximum flux of each remaining pixel time series and mask out the 10\% of pixels that reach the highest flux values. Removal of the brightest pixels eliminates sequences with the highest Poisson noise. A retained pixel will strengthen a detected signal only if the flux observed from the solar system object of interest is greater than the Poisson shot noise from the flux counts within that pixel. 

The bright eliminated pixels are typically associated with stars and bright asteroids in the frame that would reduce the signal-to-noise ratio of any detections. If an object passes over one of these masked pixels, it makes no contribution to the signal determined across the full shift-stack. As a result, if a TNO lies directly on top of masked pixels during the majority of a TESS sector, it will likely not be detected by our algorithm. To be detectable, a TNO must cross over masked pixels for a sufficiently small fraction of a TESS sector -- defined such that the total summed flux over the unmasked frames produces a $\geq5\sigma$ signal above zero flux in our aggregate frame.

\vspace{5mm}
\subsubsection{Polynomial Baseline Subtraction}
\label{section:poly_baseline_sub}
Our first baseline subtraction method uses a low-order polynomial to fit and subsequently subtract off the baseline of each pixel, with flux as a function of time $F(t)$ given by

\begin{equation}
    F(t) = k_0 + k_1 t + k_2 t^2 + ... + k_n t^n.
\end{equation} 
Here, $k_n$ are constants with values determined in the fitting process. We fit the two halves of the light curve separately and consider polynomials with degree $n_p = 1-5$ for each, calculating the reduced $\chi^2$ value,

\begin{figure}
    \centering
    \includegraphics[width=0.98\linewidth]{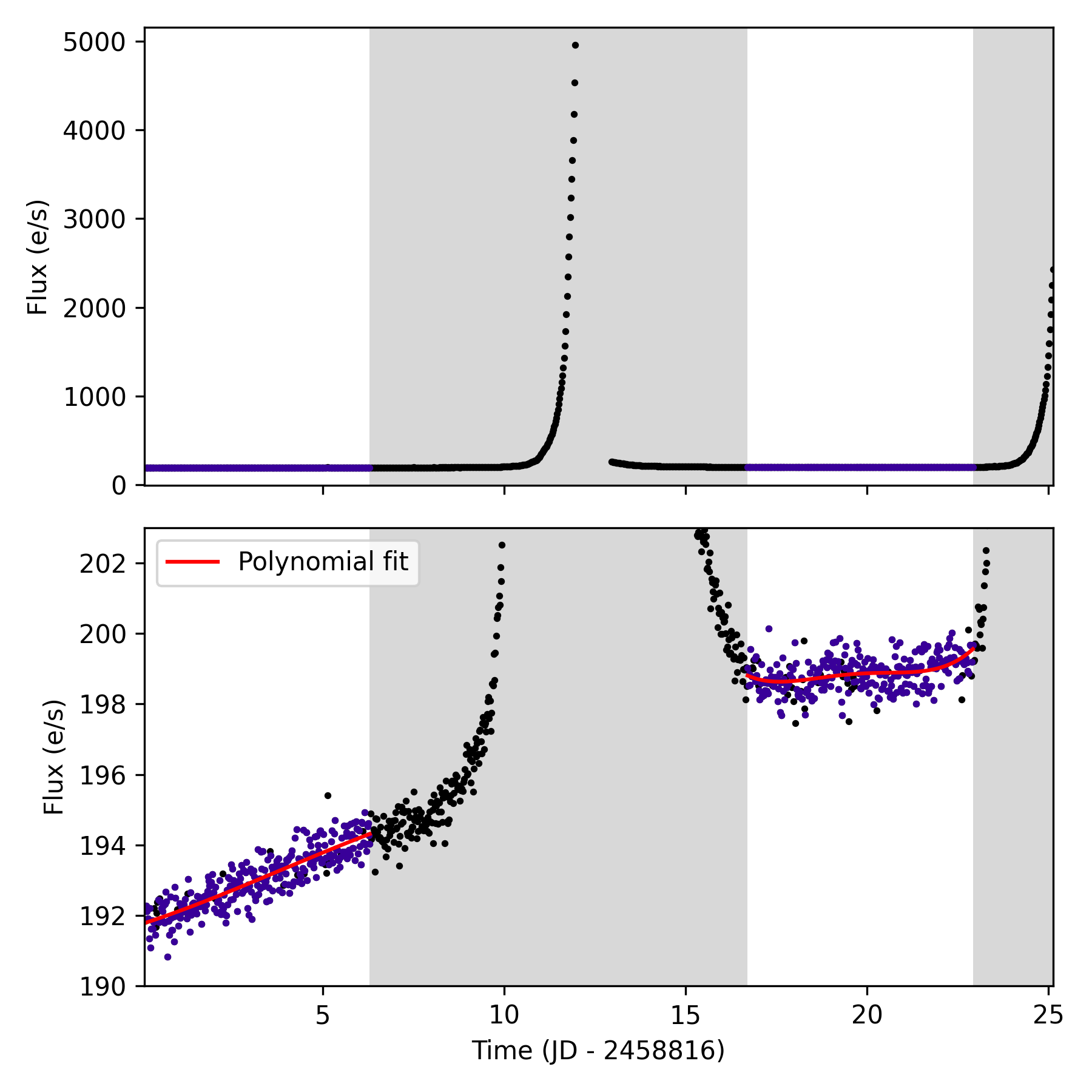}
    \caption{Sample polynomial fits to the light curve of a representative Sector 19 pixel. The top panel shows the full flux range covered by the pixel's time series, while the bottom panel zooms in to the regions that we fit after discarding the sloping systematics dominant at the middle and end of the time series. The gray shaded regions are excluded from our analysis, and times included in the baseline fit and subsequent shift-stack are shown in purple. 540 frames are retained after masking. Each light curve component is fit separately with a polynomial shown in red.}
    \label{fig:poly_bkgsub_demonstration}
\end{figure}

\begin{equation}
    \chi^2 = \frac{1}{(n_t - n_{p})}\sum_{i=1}^{i=n_t}{\Big(\frac{(F_{\mathrm{obs,} i} - F_{\mathrm{fit,} i})^2}{\sigma_i^2}\Big)}
\label{eq:reduced_chisq}
\end{equation}
for each fit. The total number of frames in the time series is given by $n_t$, whereas $F_{\mathrm{obs,} i}$ and $F_{\mathrm{fit,} i}$ are the observed and polynomial fit flux values, respectively, at frame $i$, and $\sigma_i$ is the uncertainty reported for the fitted pixel at frame $i$ in the full-frame image FITS file. We iterate over $n_p$ values for each pixel, keeping only the fit that results in the lowest $\chi^2$ value from Equation \ref{eq:reduced_chisq}. We independently apply this fitting procedure to both light curve segments associated with each pixel. An example fit for a Sector 19 light curve is shown in Figure \ref{fig:poly_bkgsub_demonstration}.

This baseline subtraction procedure makes no attempt to preserve information regarding the baseline flux of each pixel relative to its neighbors. As a result, if any pixels in the frame still include substantial stellar flux after the removal of bright pixels described in Section \ref{section:prefit_pix_mask}, the baseline flux of the star(s) in the pixel should also be removed by this process. This helps to prevent spurious ``signals" that appear from stacking frames along paths that cross over the locations of stars that have not been fully masked. The polynomial baseline subtraction method assumes a smooth flux profile that can be well-captured by a polynomial function, so it performs more poorly for pixels with substantial variability -- for example, those containing a transient source with strong flux variability over a time scale of $\sim$days.

\subsubsection{PCA Baseline Subtraction}
\label{section:pca_baseline_sub}

In our second baseline subtraction method, we utilize the \texttt{RegressionCorrector} class\footnote{See \url{https://docs.lightkurve.org/api/lightkurve.correctors.RegressionCorrector.html} for documentation.} of the \texttt{lightkurve} Python package to estimate the baseline flux profile of each pixel. This method reduces the properties of $N$ regressors, or surrounding pixels, into their constituent components using Principal Component Analysis. Reduction to these principal components removes long-term variability and stochastic noise from the estimated baseline. The PCA components are then combined with the \texttt{RegressionCorrector} to determine the best-fitting baseline correction using linear regression.

For each pixel subtraction, we use the 2000 nearest pixels as regressors. We exclude from this analysis all pixels that lie within 5 pixels of the pixel for which the baseline is being determined. In this way, we ensure that light from an object of interest is not incorporated into its baseline subtraction, reducing the risk of self-subtraction. We use three PCA components; from initial testing, we find that additional components add to the algorithm's computation time without a substantial improvement in performance.

The PCA method described here determines the baseline flux of a pixel based on its neighbors in the frame, meaning that residual long-term signals that remain after our initial masking are not removed by this baseline subtraction. While this reduces the likelihood of self-subtraction for slowly-moving objects such as Planet Nine, it also increases the rate of false positives due to an incomplete subtraction of stellar signals. For this reason, we require that all candidates are first detected using the polynomial baseline subtraction method before verifying their recoverability with the PCA method.

\subsection{Shift-Stacking}
\subsubsection{Description of the Algorithm}

After completing the baseline subtraction process, we input the reduced images into our shift-stacking algorithm to recover TNO candidates of interest. Our pipeline operates in two modes: one designed to recover known objects along a known path, and another designed to implement a blind search for unknown objects.

\begin{figure}
    \centering
    \includegraphics[width=0.48\textwidth]{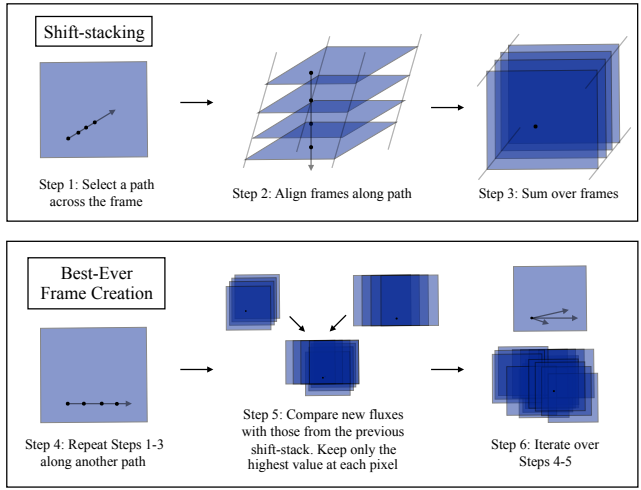}
    \caption{Overview of the shift-stacking algorithms included within our pipeline. The top row (Steps $1-3$) describes the general process of shift-stacking and can be applied to recover known objects along their projected orbits. The bottom row (Steps $4-6$) illustrates the additional steps implemented to create the ``best-ever" frames used in our blind search.}
    \label{fig:shift_stack_schematic}
\end{figure}

When we have prior knowledge of a TNO's path, it is possible to recover the object's signal by simply co-adding all baseline-subtracted frames along the known path using the process outlined in Steps $1-3$ of Figure \ref{fig:shift_stack_schematic}. In this mode of operation, our pipeline collects and sums the small, systematic flux increase from the TNO over the full time series, resulting in a point-like signal. By contrast, fluctuations in the surrounding pixel values are produced by Poisson noise; they should, as a result, sum to values centered on zero that adhere to $\sqrt{N}$ statistics. We demonstrate in Section \ref{subsection:recovery_known_objects} the results of this method as applied to several known objects in the TESS dataset.

The discovery of objects with \textit{unknown} locations and orbital parameters is a more challenging task. To find new candidate objects in trans-Neptunian space, we shift and stack our baseline-subtracted images across all plausible orbital trajectories to create an aggregate ``best-ever" frame using the process described in Steps $4-6$ of Figure \ref{fig:shift_stack_schematic}. These best-ever frames collate the results from all shift-stack paths of interest in a region of the sky into a single, information-dense aggregate frame.

We initiate the creation of a best-ever frame by placing the first baseline-subtracted image of our time series in the center of an empty image -- a 2D array of zeroes. As illustrated in Figure \ref{fig:shift_stack_schematic}, we then shift and co-add all images in the time series along a path of interest. Because our baseline-subtracted images record only deviations from the flux baseline, each constituent pixel stack, in the absence of complex systematics or objects in the frame, sums to a small number close to zero.

We then compare our post-shift-stacking image to the previous image and save only the highest flux value attained by each pixel (Step 5 in Figure \ref{fig:shift_stack_schematic}). For the first shift-stack path, all positive values are saved, since our original image consists of only zeroes. We repeat this process along all possible paths; after many iterations, the zeroes of the original image are replaced with positive values. The emergent best-ever frame tracks systematic positive deviations from the baseline flux along a path. We stress that these frames are not sky images, but, rather, a graphic assessment of all the shift-stack paths under consideration within a single patch of sky.

We make some simplifying assumptions to minimize the computational expense of this search. First, we assume a constant speed across the frame for each TNO and do not incorporate deviations from a linear path between the start and end pixel. This approximation should hold for all bound TNOs of interest, since their orbital motion generates only relatively small deviations from a primary projected path determined by Earth's parallax. We also consider only integer total pixel shifts, and we shift and stack frames only at integer pixel intervals. While our pipeline includes the option to interpolate to sub-pixel shifts, we found that this alteration significantly slows down the pipeline without substantially changing the results. Sub-pixel shifts may be useful in future iterations of this work to maximize the recovered flux from a source; this could be especially useful near the edges of our current detection limits.

Without interpolation, the shift-stacking component of our pipeline consists of only quick array operations set up in a grid, with $(n_x + 1) \cdot (n_y + 1)$ operations for $n_x$ $x-$shifts and $n_y$ $y-$shifts. For example, for $4 < \Delta x <  47$ pixel shifts in the $x-$direction and  $-8 < \Delta y <  8$ pixel shifts in the $y-$direction, used in our blind Sector 18 search, $n_x=43$ and $n_y=16$ for a total of 748 array operations.

The range of orbits considered in a given search determines the number of shifts necessary to include in this grid. In the parameter space explored by this work, where Earth's motion dominates objects' motion along the detector, the range of distances spanned by the population of interest at the observing epoch is the primary determining factor for appropriate values of $n_x$ and $n_y$. This distance, together with the projected magnitude of an object, is also our primary observable for the distant solar system objects studied in this survey, which subtend short orbital arcs spanning consecutive nights (see e.g. \citet{bernstein2000orbit}). We refer the reader to Section \ref{subsection:orbit_interpretation} of this manuscript for a description of the relationship between an object's orbital elements and its projected path on the TESS detector, which can be used to deduce the $x-$ and $y-$shifts of interest.

\subsubsection{Distinguishing between systematic and true signals}
\label{subsubsection:systematics}

The excess flux from an astronomical object is maximized by summing along the closest possible path to that of the object's true orbit. Similar paths that only partially align with the correct orbit produce an excess in flux, but that excess is weaker than that from the shift-stack that most closely fits the object's track. Thus, a properly recovered object should appear as a point source with a surrounding ``cloud" of radially decreasing brightness, where the central pixel has the peak flux and the ``cloud" corresponds to similar orbit paths that overlap with, but are slightly misaligned with, the true path.

Imperfections in the baseline subtraction method can also lead to artificial flux excesses. If the determined baseline for a given pixel does not closely follow that pixel's true profile, a bias is introduced such that the baseline-subtracted light curve no longer follows Poisson statistics about a baseline of zero. The bias is then propagated forwards such that the associated pixel stacks no longer sum to values centered around zero, producing systematic features in our best-ever frames.

Bright asteroids outside of our search limits serve as a major source of systematics in the shift-stacked frames. Asteroids moving across the frame can produce strong spikes in individual pixel light curves, leading to a peaked flux profile that is captured by neither a polynomial fit nor a PCA fit using the surrounding pixels. Our initial pixel masking process removes only gradients that are uniformly present across the entire frame, rather than those attained by individual pixels.

Fortunately, these systematics typically manifest as extended structures in the best-ever frames, without the tapering ``cloud"-like effect of the true objects (see Section \ref{subsection:recovery_known_objects} for specific examples). Thus, differentiating between systematics and true objects becomes a challenging task only for very dim objects or very slow-moving objects (see Section \ref{subsection:injection_recovery}). We apply both baseline subtraction methods, which each produce different systematic patterns, to mitigate this problem.

\subsection{Automated Candidate Extraction}
After creating our best-ever frames(Steps 4-6 in Figure \ref{fig:shift_stack_schematic}), we then identify all local maxima in these frames with summed flux $\geq3\sigma$ above zero, where the standard deviation is computed across the entire frame. We accomplish this by first applying a maximum and minimum filter to the best-ever image, each with a neighborhood size of 5 pixels. We then find locations where the difference between the maximum and minimum filtered images is at least $3\sigma$. To quickly identify the strongest signals, we distinguish between sources recovered at $3-5\sigma$ and at significance higher than $5\sigma$. Finally, we calculate the weighted center of flux of each identified local maximum to more precisely pinpoint each candidate location in the frame. 

By applying this procedure to the best-ever frames, rather than to each individual shift-stacked frame, we simultaneously extract all of the strongest signals in a given frame. Our significance thresholds are set relative to zero, meaning that they are more representative of an object's significance in a single shift-stack frame (obtained from Steps 1-3 in Figure \ref{fig:shift_stack_schematic}) than its significance relative to other shift-stacking combinations of neighboring pixels. However, we note that the standard deviation of the best-ever frames may substantially differ from that of an individual shift-stack frame. 

In its current form, our automated source extraction algorithm does not distinguish between point-like sources and more extended sources. As a result, all sources must be further examined to verify whether they resemble compact signals rather than elongated systematics. Regardless, the automated algorithm serves as a useful tool to quickly identify possible sources of interest and to guide the eye. Future developments of this pipeline will replace this module with an alternative computer vision algorithm to efficiently distinguish between systematics and true signals in a more fully automated manner.

\subsection{Full Pipeline Workflow - Blind Candidate Search}
Combining our three pipeline components, we iteratively run the full pipeline on each frame cutout across each camera and CCD of a sector. After initial pixel masking -- both in the temporal and spatial dimensions -- we apply our polynomial baseline subtraction to individual cutout regions. We shift-stack the reduced images, then use the results to select promising sources with our automated candidate extraction algorithm. We vet these candidates by eye and select only those that resemble point-like signals. Then, we re-reduce the cutout regions with identified promising sources using the more computationally expensive PCA baseline subtraction. We repeat the shift-stack process and the automated candidate extraction, then cross-match to find which candidates were re-recovered.

\section{Results}
\label{section:results}

After developing our pipeline, we demonstrated its performance by recovering the signals of three known outer solar system objects. We then applied the same framework to blindly search for new candidate objects in TESS sectors 18 and 19, using injection tests to verify the performance of our pipeline. Finally, we developed a formalism to interpret the orbits of objects recovered in the TESS frames with shift-stacking.

\subsection{Search Limits}
Because outer solar system bodies have slow orbital velocities relative to that of Earth, their movement across the TESS CCDs is dominated by Earth's parallactic motion. All of the TESS cameras are aligned with the ecliptic plane in which Earth orbits, meaning that, over a 27-day observing sector, slowly-orbiting outer solar system objects primarily move in one direction -- horizontally across the TESS frame (in the $x$-direction) -- with little vertical ($y$-direction) motion. As a result, we set the $x$-pixel shift limits of our searches based on the expected parallactic movement of an object on a circular orbit at the distance of interest (see Section \ref{subsection:orbit_interpretation}). We also allow for shifts of up to $\pm8$ pixels ($\pm168\arcsec$) in the $y$-direction across the masked temporal baseline to account for orbital motion on high-inclination orbits.

The TESS spacecraft itself is not stationary relative to the Earth; it follows a highly eccentric geocentric orbit -- characterized by perigee and apogee at $17R_{\oplus}$ and $59R_{\oplus}$, respectively -- with a 13.7-day period in a 2:1 resonance with the Earth's moon \citep{ricker2015tess}. For an object at $d=35$ au, the difference between perigee and apogee can induce a positional shift of up to $19\arcsec$ ($<1$ pixel) on the TESS detector. While this sub-pixel shift is too small to affect the results presented here, which focus on the distant solar system ($d\geq 35$ au), the TESS spacecraft orbit should be taken into account in studies focusing on more nearby solar system populations -- particularly objects interior to $d=32$ au, where the TESS spacecraft motion can induce a shift exceeding 1 pixel.

\subsection{Recovery of Known Objects}
\label{subsection:recovery_known_objects}
\begin{figure*}
    \centering
    \includegraphics[width=0.98\linewidth]{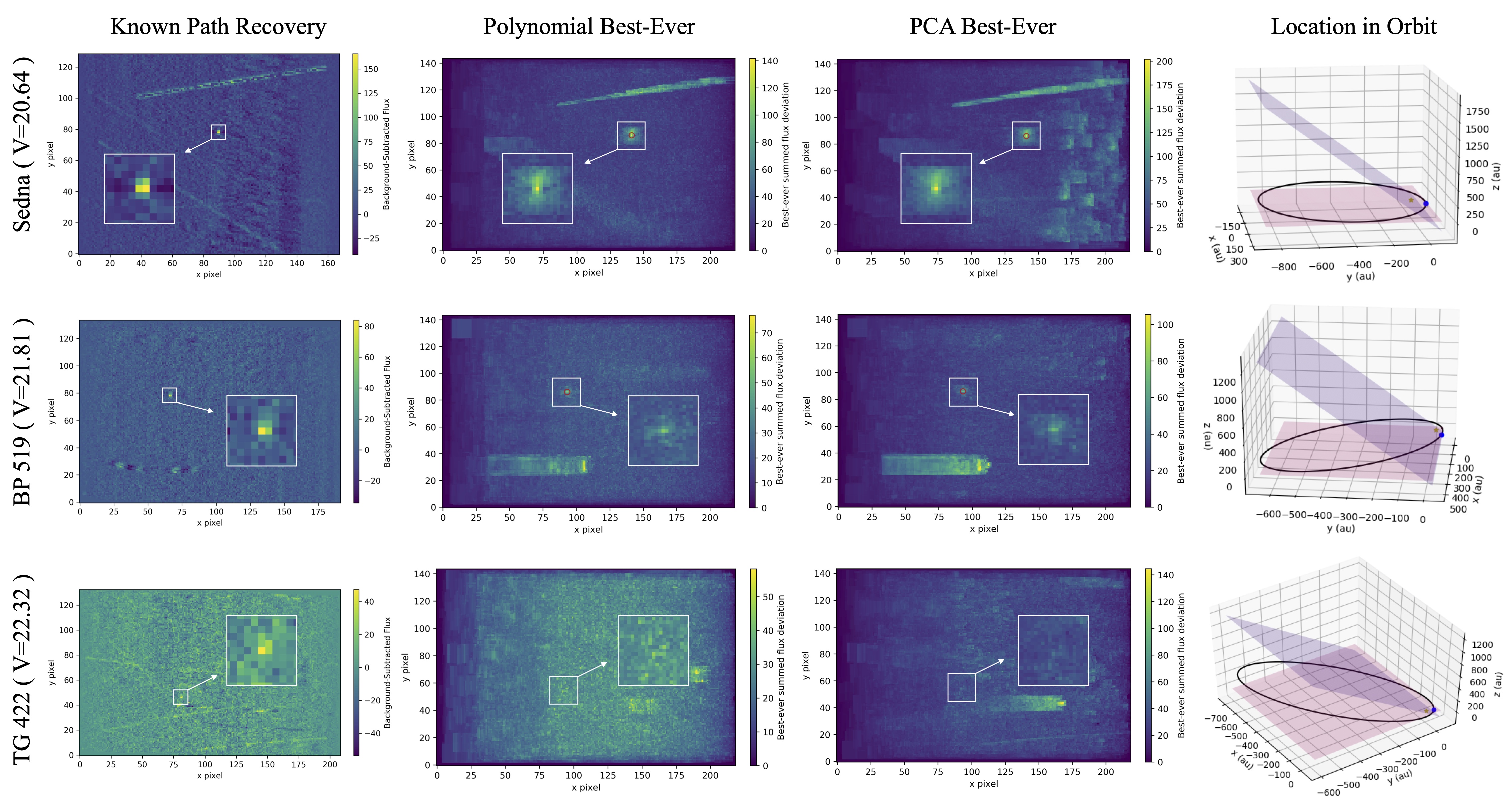}
    \caption{Dim solar system objects Sedna ($V=20.64$), 2015 BP519 ($V=21.81$), and 2007 TG422 ($V=22.32$) recovered through our shift-stacking pipeline. The first column displays the PCA baseline-subtracted recovery of each object along its known projected path on the sky, with varying image dimensions set by the path length covered by the object. The second and third columns show best-ever frames of the same objects, which compile the results of a blind search for any objects with $x$-pixel shifts corresponding to $d=35-800$ au and with $y$-pixel shifts $-8 < \Delta y < +8$ across the single-sector baseline of TESS in a $128 \times 128$ pixel frame encompassing the known objects. No prior orbital information about the objects is incorporated in these blind recoveries. The fourth column shows the location of each object in its orbit at the time of recovery, with the ecliptic plane shaded in red and the galactic plane shaded in blue for reference.}
    \label{fig:recovered_obj_fig}
\end{figure*}

We begin by testing the pipeline's ability to blindly recover the known outer solar system objects listed in Table \ref{tab:known_obj_table}. We show that distant TNOs with magnitudes down to $V \sim 22$ are readily recoverable and distinguishable from systematic effects in our Sector 5 best-ever frames, and we provide three examples of known TNOs -- 90377 Sedna \citep{brown2004discovery}, 2015 BP519 \citep{becker2018}, and 2007 TG422 -- run through our pipeline using its two modes of operation: with input orbital information and with our blind search method. To encompass the orbits of each of these objects, our blind searches in this section span pixel shifts corresponding to circular orbits at distances between 35 and 800 au. Our results are summarized in Figure \ref{fig:recovered_obj_fig}, and we describe each individual recovery below.

\begin{deluxetable*}{cccccccccccc}
\tablecaption{Blind recovery results for the three known objects shown in Figure \ref{fig:recovered_obj_fig}. Values are reported at the last unmasked time in Sector 5, at $t=$2458461.19 JD (December 8, 2018) for all three frames. Nominal values were extracted from JPL Horizons at the same epoch, and radii were computed using $p_V=0.32, 0.08$ and $0.04$ for Sedna, 2015 BP519, and 2007 TG422, respectively. We do not include nominal shift-stack paths for these objects because, while the projected path of each object is known, nonlinearities in these paths imply that the ``best" recovery may not be a straight line from the start to end location of the object during this time series. The cutout origin is reported in pixels, referenced from the first data pixel of the FFI. 2015 BP519 also crosses through Sector 4, Camera 3, CCD 1; for a direct comparison with the other two objects, we elect to include only its Sector 5 track in our analysis. Because 2007 TG422 was not recovered in the corresponding best-ever frame, we include only its nominal and known-path recovery values here for reference. \label{tab:known_obj_table}}
\tablewidth{700pt}
\tabletypesize{\scriptsize}
\tablehead{
\colhead{Name} & \colhead{(S, Cam, CCD)} & \colhead{Cutout Origin} & \colhead{Type} & \colhead{RA (deg)} & \colhead{Dec (deg)} & \colhead{($\Delta x, \Delta y$)} & \colhead{$V$} & \colhead{$d$ (au)} & \colhead{$r$ (km)} & \colhead{$N_{\rm frames}$} & \colhead{Significance}}
\startdata
90377 Sedna & (5, 1, 4) & (1450, 1024) & nominal & 56.8960 & 7.6094 & - & 20.64 & 86.2 & 500 & - & - \\
& & & PCA, known & - & - & - & - & - & - & 733 & 15.3$\sigma$ \\
& & & poly, blind & 57.1268 & 7.6637 & (40, -2) & 20.81 & 93.3 & 539 & 733 & 11.6$\sigma$ \\
& & & PCA, blind & 57.1268 & 7.6637 & (40, -2) & 20.43 & 93.3 & 644 & 733 & 8.7$\sigma$ \\ \hline
2015 BP519 & (5, 3, 2) & (900, 70) & nominal & 66.8552 & -33.6904 & - & 21.81 & 54.4 & 299 & - & - \\ 
& & & PCA, known & - & - & - & - & - & - & 733 & 14.3$\sigma$ \\
 & & & poly, blind & 67.3436 & -33.3435 & (68, 6) & 22.25 & 54.8 & 95.9 & 733 & 3.2$\sigma$\\
 & & & PCA, blind & 67.1341 & -33.6660 & (65, -3) & 21.55 & 57.5 & 146 & 733 & 5.5$\sigma$ \\ \hline
2007 TG422 & (5, 1, 3) & (1367, 1260) & nominal & 69.9593 & 8.7433 & - & 22.32 & 36.8 & 168 & - & - \\
& & & PCA, known & - & - & - & - & - & - & 736 & 5.8$\sigma$ \\
\hline
\enddata
\end{deluxetable*}

\subsubsection{90377 Sedna}
One particularly interesting test case is that of Sedna, a detached Kuiper belt object described in the first row of Table \ref{tab:known_obj_table}. Sedna has a high eccentricity ($e=0.86$) and large semimajor axis ($a=484$ au), and its current proximity to perihelion in its orbit ($q=76.1$ au) makes it one of the brightest detached KBOs, at magnitude $V = 20.64$ during the time of the TESS Sector 5 observations. Sedna's location in its orbit at the time of observation is shown in the top right panel of Figure \ref{fig:recovered_obj_fig}, with the ecliptic and galactic planes included for reference. Sedna is a prime test object for our algorithm, since it is a distant object with a correspondingly short path length across the TESS cameras (42 pixels over the 22-day baseline of Sector 5 data kept after initial masking) which lies below the TESS single-frame detection limit.

We illustrate our known path recovery of Sedna in the top left frame of Figure \ref{fig:recovered_obj_fig}, where we recovered Sedna's signal at $15.3\sigma$. We define the significance of the known-path recoveries relative to the standard deviation of the full frame after masking out the $8\times8$ pixel region surrounding the recovered object. 

This recovery was obtained by summing 733 PCA baseline-subtracted FFIs along Sedna's known path. Because we complete the shift-stacking process without accounting for Sedna's sub-pixel location, the primary flux peak is shared among a few neighboring pixels. This is expected, since a moving object spends the most time towards the edges of a pixel, resulting in substantial flux spillover into neighboring pixels. Furthermore, because of the TESS PSF, only 50\% of the flux from an object falls within 1 pix$^2$ of the object's peak flux location.

The second and third columns of Figure \ref{fig:recovered_obj_fig} show our pipeline's blind recovery results obtained using the polynomial and PCA baseline subtractions, respectively. We emphasize that these two frames are both best-ever images obtained with no prior information related to Sedna's location or expected properties. Despite the absence of this input information, we recover Sedna at high significance in both images: at 11.6$\sigma$ in the polynomial recovery and at 8.7$\sigma$ in the PCA recovery. In the best-ever images, the standard deviation used to set the recovery significance is determined after masking out the $20\times20$ pixel region surrounding the recovered object, since the recovered signals are substantially more extended than in the known-path recoveries.

\subsubsection{2015 BP519}
2015 BP519 is a high-eccentricity ($e=0.923$) extreme trans-Neptunian object with semimajor axis $a = 454$ au and perihelion $q=35.2$ au. For assumed albedo $p_V=0.08$ corresponding to a typical red TNO  \citep{stansberry2008physical}, 2015 BP519 has radius $r=299$ km. The object's known-path recovery, best-ever recoveries, and location along its orbit are portrayed in the second row of Figure \ref{fig:recovered_obj_fig}. Each of the stacked 2015 BP519 frames in Figure \ref{fig:recovered_obj_fig} aggregates the signals of 733 TESS FFIs after initial masking.

2015 BP519 is readily recoverable through shift-stacking due to its current location near perihelion and its relatively large radius, though its fainter magnitude results in a weaker recovery than that of Sedna. The recovery of 2015 BP519 along its known path is only marginally weaker than the corresponding recovery for Sedna, with a $14.3\sigma$ peak. The best-ever frames, on the other hand, show substantially weaker detections than those of Sedna, likely owing to the higher magnitude of 2015 BP519. While 2015 BP519 still produces a clear signal recovered at high significance, its weaker recoveries suggest that it is approaching the magnitude limit of our blind search.

\subsubsection{2007 TG422}
2007 TG422 is the dimmest of the three sample TNOs that we recover, and it clearly demonstrates the power of shift-stacking to recover even very dim ($V\sim22.3$) signals with confidence. Though it is the nearest of the three objects detailed in Table \ref{tab:known_obj_table}, at a distance $d=36.8$ au, 2007 TG422 is also significantly smaller than Sedna and 2015 BP519, with radius $r=168$ km assuming $p_V=0.04$ -- appropriate for a neutral-colored TNO such as 2007 TG422 \citep{stansberry2008physical}. With eccentricity $e=0.931$, semimajor axis $a=512$ au, and perihelion $q=35.6$ au, 2007 TG422 is currently observable due to its proximity to perihelion, as shown in the bottom right panel of Figure \ref{fig:recovered_obj_fig}.

Our $5.8\sigma$ known-path recovery of 2007 TG422, aggregated over 736 frames, corresponds to a contributed flux of only 0.0645 e/s, well below the single-frame detection limit of TESS. As a result, the known path recovery of 2007 TG422 produces a point source signal just marginally brighter than the background. In both of the two best-ever frames, no local maximum is found at the expected location of 2007 TG422. This indicates that the TNO's signal is not strong enough to be recovered by our blind search, placing a clear bound on our magnitude limit.

\subsubsection{Systematics and Uncertainties}
In addition to the signals of interest, each of the frames in Figure \ref{fig:recovered_obj_fig} also includes systematic features. Systematics persist due to imperfections in the baseline subtraction process, leading to residual effects that are asymmetric about the flux baseline.

\begin{figure*}
    \centering
    \includegraphics[width=0.98\linewidth]{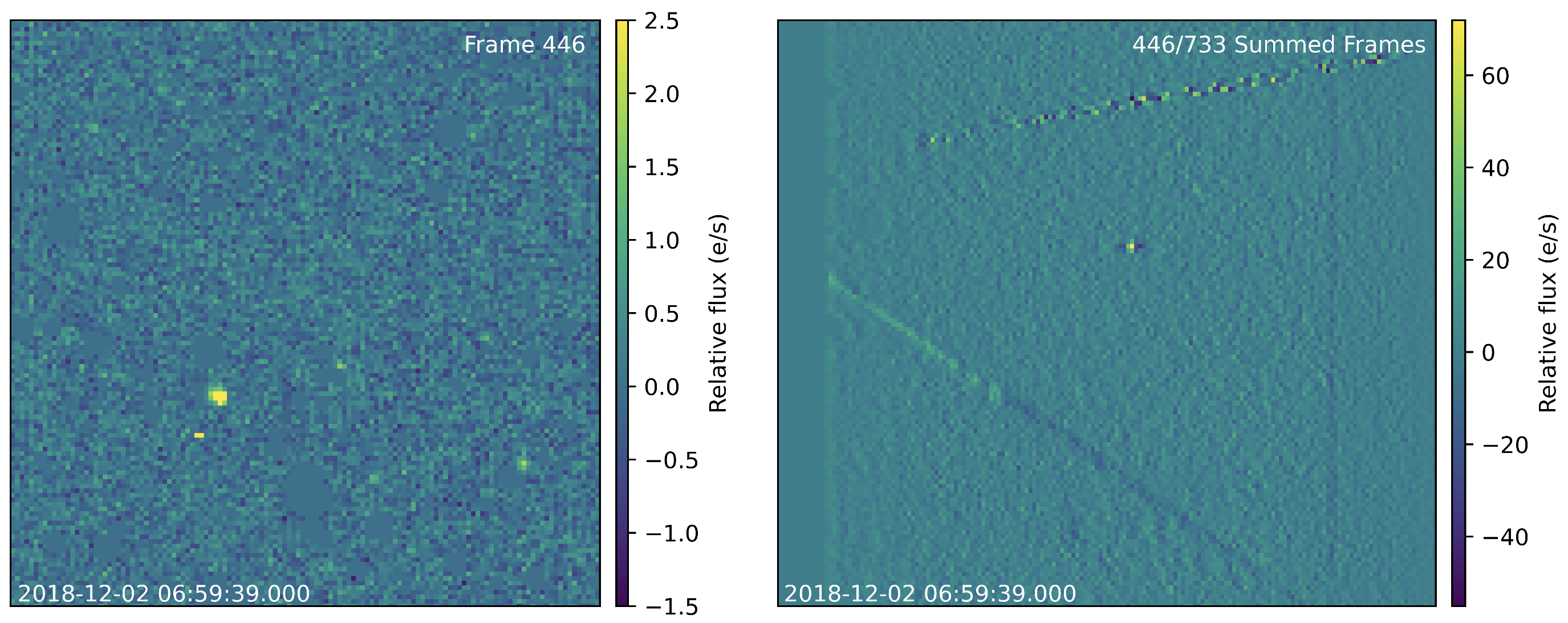}
    \caption{Animation of all 733 polynomial baseline-subtracted frames used to recover Sedna in Figure \ref{fig:recovered_obj_fig}. The paths of two bright asteroids passing through the frame from 2018-11-18 to 2018-11-21 and from 2018-12-01 to 2018-12-04 produce diagonal systematics in the corresponding Sedna best-ever frame.}
    \label{fig:animated_sedna}
\end{figure*}

Each recovery of Sedna in Figure \ref{fig:recovered_obj_fig} includes a long, diagonal streak towards the top of the frame. Figure \ref{fig:animated_sedna} provides intuition for the origin of this systematic feature, which directly corresponds to a bright asteroid passing over the frame (see the frames spanning 2018-11-18 to 2018-11-21). Though the central pixels along the asteroid's path were masked due to their high maximum flux, the surrounding regions remain unmasked. This results in a high transient flux within pixels coincident with the asteroid's path, leading to a high summed flux when those pixels are included in a shift-stack.

The path of a second bright asteroid (see 2018-12-01 to 2018-12-04 in Figure 5) is also evident as a diagonal streak in the lower region of Sedna's known path recovery, while the corresponding feature is more diffuse in the best-ever frames. The diffusion of the second asteroid's signal is likely due to the larger y-component of its movement, meaning that the blind shift-stack range of interest does not capture movement directly along the asteroid's orbit.

Systematics produced by foreground asteroids should be most common close to the ecliptic plane of the solar system, corresponding to TESS Camera 1 (where Sedna is located). While we made no direct effort to discard these signals in this work, the automated removal of such signals can be accomplished through cross-matching with existing asteroid catalogs to further clean the TESS dataset prior to larger-scale surveys.

In the PCA best-ever frame, numerous block-like systematics are present to the right of Sedna, while these systematics are absent in the polynomial subtraction frame. These features likely result from an incomplete removal of residual stellar fluxes in the PCA baseline subtraction, which does not incorporate the flux baseline of the pixel at hand. We find from this example and others in our blind search that the PCA best-ever frames tend to be noisier and thus more prone to false positive signals than the polynomial-subtracted frames. However, as we describe in Section \ref{subsection:injection_recovery}, the PCA frames are also more sensitive to dim, slow-moving signals and thus contribute great value to the Planet Nine search.

Horizontally extended systematic signals are present in the 2015 BP519 and 2007 TG422 best-ever frames, as well. These systematics are not directly attributable to asteroids, and they appear with higher frequency near the galactic plane. This suggests that they may be associated with stars that have not been fully masked from the frame.

The maximum fluxes attained in each of the PCA best-ever images are notably higher than those of their polynomial best-ever counterparts. This is likely because the polynomial subtraction method is more prone to self-subtraction, since the baseline profile of each pixel is determined solely from that pixel's time series flux, rather than surrounding ``background" pixels in the frame. The known path recovery frame peaks at a higher flux than the polynomial best-ever frame, meaning that the path determined from our best-ever frame is similar to, but not quite the same as, Sedna's actual path. Deviations between these two maximum fluxes stem from nonlinearities in the object's real path across the TESS frame, which are ignored in the best-ever image creation. 

While our PCA best-ever frames can be used to circumvent the issue of self-subtraction, uncertainties related to the extent of these path nonlinearities are inherent to our linear search method. As a result, our magnitude estimates are systematically lower than the real object magnitudes, and this effect is more apparent for more nearby objects with stronger nonlinearities. The path uncertainty and thus dilution of signal reduces the range of magnitudes recoverable with this method; however, it also suggests that recovered objects should be more amenable to follow-up observations, since their actual magnitudes may be slightly elevated over those suggested by our best-ever frames.

\subsection{New Candidates from a Blind Search}
\label{subsection:new_candidates}

\subsubsection{Candidate Extraction}
After verifying our ability to recover known objects, we then conducted a blind search for previously unknown outer solar system objects. We completed this search using all FFIs from Sectors 18 and 19. Due to the presence of the bright star Polaris \citep[$I = 1.22$;][]{ducati2002vizier} in Camera 3, CCD 3 of Sector 19, a column of pixels across the CCD was saturated, leading to an overestimated smear correction affecting multiple columns of pixels.\footnote{\url{https://archive.stsci.edu/missions/tess/doc/tess_drn/tess_sector_19_drn26_v02.pdf}} These columns produced strong systematics in our pipeline; as a result, we discarded the affected $256\times256$ frames, which constitute a small fraction of our total sample.

We searched all paths with $x$-shift bounds set to match $a=70-800$ au for circular orbits: around $4 \leq \Delta x < 47$ pixels after the initial masking step has reduced the full temporal baseline to $\sim$19 days in Sector 18, and around $5 \leq \Delta x < 58$ pixels for the $\sim23$-day baseline in Sector 19. As in Section \ref{subsection:recovery_known_objects}, we also allowed for $y$-shifts of up to $\pm8$ pixels across the masked baseline to account for small path deviations from solely parallax-dominated motion. After initial masking, our final image stacks included 574 frames in our Sector 18 analysis and 540 frames in our Sector 19 analysis. This is roughly 25\% fewer frames than were used in the Sector 5 recoveries described in Section \ref{subsection:recovery_known_objects}, meaning that our magnitude limit will accordingly be about 0.3 mag higher before accounting for the increased stellar density in the galactic plane.

Our automated extraction process returned a large number of $\geq3\sigma$ local maxima (typically between a few and a few tens per $256\times256$ frame) from the polynomial baseline-subtracted best-ever frames. However, upon visual inspection, we found that most of these local maxima could be quickly discarded as systematics. We carefully inspected all frames, taking note of those containing potentially promising candidate sources, and re-ran the promising frames using the more computationally intensive PCA baseline subtraction method.

We cross-compared frames to determine whether the point-like features present in the polynomial-subtracted best-ever frames were recovered with the PCA subtraction. In many cases, the second baseline subtraction revealed that these features more closely resembled systematics when processed in a different manner. By comparing our results from both reductions, we narrowed down our set of candidates to those listed in Table \ref{tab:candidate_list_table}. 

\begin{deluxetable*}{ccccccccccccc}
\tablecaption{Candidates recovered in best-ever frames obtained with both baseline subtraction algorithms. We report values recovered from both subtraction methods. Coordinates are reported at the last unmasked time in the sector, and the reported distances ($d$) refer to the predicted distance between the candidate object and the TESS spacecraft at the epoch of detection. For objects recovered in two separate stacks, four entries are included in the table, with results from the second stack provided as the third and fourth rows. Significances are reported as the deviation above zero flux recovered in our automated candidate extraction, where the standard deviation is calculated across the full best-ever frame.} \label{tab:candidate_list_table}
\tablewidth{700pt}
\tabletypesize{\scriptsize}
\tablehead{
\colhead{$N_{\rm cand}$} & \colhead{(S, Cam, CCD)} & \colhead{Cutout Origin} & \colhead{Type} & \colhead{RA (deg)} & \colhead{Dec (deg)} & \colhead{($\Delta x, \Delta y$)} & \colhead{$V$} & \colhead{$d$ (au)} & \colhead{$r$ (km)} & \colhead{$N_{\rm frames}$} & \colhead{Epoch (JD)} & \colhead{Significance}}
\startdata
1 & (18, 2, 1) & (256, 384) & poly & 43.9497 & 69.3189 & (18, 1) & 21.11 & 167.7 & 1517 & 573 & 2458810.25 & 9.29$\sigma$ \\
 & & & PCA & 43.9497 & 69.3189 & (18, 1) & 21.15 & 167.7 & 1487 & 573 & 2458810.25 & 3.98$\sigma$ \\ \hline
2 & (18, 2, 3) & (0, 1664) & poly & 16.6250 & 49.8912 & (24, 4) & 20.80 & 123.5 & 948 & 573 & 2458810.25 & 6.69$\sigma$\\
 & & & PCA & 16.6250 & 49.8912 & (24, 4) & 20.56 & 123.5 & 1060 & 573 & 2458810.25 & 4.31$\sigma$ \\ \hline
3 & (18, 3, 3) & (1280, 1664) & poly & 344.4852 & 78.6428 & (36, -3) & 21.32 & 83.8 & 343 & 574 & 2458810.25 & 6.09$\sigma$ \\
 & & & PCA & 344.5048 & 78.6538 & (34, -3) & 20.92 & 88.7 & 462 & 574 & 2458810.25 & 5.70$\sigma$ \\ \hline
4 & (18, 4, 2) & (1664, 1664) & poly & 244.9781 & 73.4310 & (32, 3) & 22.05 & 95.6 & 320 & 574 & 2458810.25 & 6.96$\sigma$ \\
 & & & PCA & 244.9781 & 73.4310 & (32, 3) & 21.70 & 95.6 & 375 & 574 & 2458810.25 & 5.24$\sigma$ \\
 & & (1664, 1536) & poly & 244.9781 & 73.4310 & (32, 3) & 22.04 & 95.5 & 321 & 574 & 2458810.25 & 7.20$\sigma$ \\
 & & & PCA & 244.9781 & 73.4310 & (32, 3) & 21.69 & 95.5 & 376 & 574 & 2458810.25 & 5.21$\sigma$ \\ \hline
5 & (18, 4, 2) & (768, 1664) & poly & 260.9884 & 70.8163 & (32, 0) & 22.16 & 96.3 & 309 & 574 & 2458810.25 & 5.21$\sigma$ \\
 & & & PCA & 260.9884 & 70.8163 & (32, 0) & 21.43 & 96.3 & 375 & 574 & 2458810.25 & 3.91$\sigma$ \\
 & & (768, 1792) & poly & 260.9884 & 70.8163 & (32, 0) & 22.16 & 96.3 & 310 & 574 & 2458810.25 & 5.81$\sigma$ \\
 & & & PCA & 260.9884 & 70.8163 & (32, 0) & 21.74 & 96.3 & 431 & 574 & 2458810.25 & 3.44$\sigma$ \\ \hline
6 & (18, 4, 3) & (1152, 1280) & poly & 252.6156 & 65.2268 & (38, 4) & 22.06 & 80.7 & 227 & 574 & 2458810.25 & 6.03$\sigma$ \\
 & & & PCA & 252.5973 & 65.2182 & (37, 3) & 21.4 & 83.1 & 319 & 574 & 2458810.25 & 3.91$\sigma$ \\ \hline
7 & (19, 2, 3) & (896, 1664) & poly & 57.8788 & 61.6236 & (36, 3) & 20.74 & 106.7 & 727 & 539 & 2458838.92 & 5.73$\sigma$ \\
 & & & PCA & 57.8788 & 61.6236 & (36, 3) & 20.20 & 106.7 & 933 & 539 & 2458838.92 & 3.95$\sigma$ \\ \hline
8 & (19, 3, 2) & (1664, 1024) & poly & 122.2438 & 81.4212 & (36, 0) & 22.02 & 107.1 & 407 & 539 & 2458838.92 & 5.16$\sigma$ \\
 & & & PCA & 122.2698 & 81.4253 & (36, 3) & 21.66 & 106.7 & 477 & 539 & 2458838.92 & 4.55$\sigma$ \\ \hline
9 & (19, 3, 2) & (896, 1536) & poly & 105.0573 & 86.9216 & (19, -2) & 21.77 & 201.7 & 1616 & 539 & 2458838.92 & 7.07$\sigma$ \\
 & & & PCA & 105.1514 & 86.9321 & (19, 0) & 21.62 & 202.8 & 1758 & 539 & 2458838.92 & 5.66$\sigma$ \\ \hline
10 & (19, 3, 2) & (1024, 1024) & poly & 98.0576 & 83.7762 & (42, 3) & 21.89 & 91.5 & 316 & 539 & 2458838.92 & 5.74$\sigma$ \\
 & & & PCA & 98.0063 & 83.7598 & (42, 0) & 21.48 & 91.8 & 384 & 539 & 2458838.92 & 5.02$\sigma$ \\ \hline
11 & (19, 3, 2) & (896, 1024) & poly & 99.9219 & 83.7321 & (23, 6) & 21.92 & 162.1 & 976 & 539 & 2458838.92 & 5.76$\sigma$ \\
 & & & PCA & 99.9219 & 83.7321 & (23, 6) & 21.59 & 162.1 & 1137 & 539 & 2458838.92 & 3.34$\sigma$ \\ \hline
12 & (19, 3, 2) & (1664, 1408) & poly & 132.6612 & 82.2808 & (30, 6) & 21.85 & 126.0 & 608 & 539 & 2458838.92 & 6.06$\sigma$ \\
 & & & PCA & 132.6612 & 82.2808 & (30, 6) & 21.65 & 126.0 & 667 & 539 & 2458838.92 & 4.53$\sigma$ \\ \hline
13 & (19, 3, 2) & (1408, 1664) & poly & 140.7700 & 85.2651 & (39, 5) & 21.96 & 98.0 & 350 & 540 & 2458838.92 & 5.80$\sigma$ \\
 & & & PCA & 140.7700 & 85.2651 & (39, 5) & 21.71 & 98.0 & 393 & 540 & 2458838.92 & 3.14$\sigma$ \\ \hline
14 & (19, 3, 3) & (640, 1280) & poly & 206.4893 & 85.7287 & (47, -3) & 22.23 & 81.8 & 216 & 539 & 2458838.92 & 5.46$\sigma$ \\
 & & & PCA & 206.3024 & 85.7390 & (47, 0) & 21.81 & 82.0 & 263 & 539 & 2458838.92 & 5.45$\sigma$ \\ \hline
15 & (19, 4, 1) & (1152, 1792) & poly & 283.4931 & 66.3750 & (32, 3) & 21.91 & 119.9 & 538 & 540 & 2458838.92 & 5.19$\sigma$ \\
 & & & PCA & 283.4931 & 66.3750 & (32, 3) & 21.40 & 119.9 & 679 & 540 & 2458838.92 & 3.33$\sigma$ \\ \hline
16 & (19, 4, 2) & (1408, 896) & poly & 242.2779 & 73.2150 & (42, 1) & 22.04 & 91.7 & 295 & 539 & 2458838.92 & 6.17$\sigma$ \\
 & & & PCA & 242.2779 & 73.2150 & (42, 1) & 21.62 & 91.7 & 359 & 539 & 2458838.92 & 5.22$\sigma$ \\ \hline
17 & (19, 4, 3) & (896, 1024) & poly & 253.6466 & 60.7441 & (45, -2) & 22.20 & 85.6 & 240 & 539 & 2458838.92 & 5.33$\sigma$ \\
 & & & PCA & 253.6466 & 60.7441 & (45, -2) & 21.85 & 85.6 & 282 & 539 & 2458838.92 & 3.04$\sigma$ \\ 
\enddata
\end{deluxetable*}

\subsubsection{Physical Property Estimates}

For each candidate, we estimated several physical properties -- sky coordinates, distance, radius, and magnitude. We extracted each property using both baseline subtraction methods, resulting in two separate estimates of each parameter in Table \ref{tab:candidate_list_table}. The discrepancy between results from each reduction method provides a sense of the parameter uncertainty for individual candidates. Below, we detail our methods for determining each of these properties.

\paragraph{Sky Coordinates}
We directly extracted the coordinates of candidates at each time based on their pixel locations in the TESS frames. Because we did not implement our shift-stacking method with sub-pixel precision, an uncertainty floor is set by the size of the pixels ($21\arcsec \times 21 \arcsec$). Our true uncertainty is higher -- on the order of a few pixels -- because only most, but not necessarily all, of the true object path is required to match the simplified shift-stacked path for our algorithm to return a recovery.

\paragraph{Distance}
The distance to each object was determined from the shift-stack path length covered by the object over the TESS observing baseline, where we assumed that the movement of an object across the sky was dominated by Earth's parallactic motion. Thus, we inferred the distance to an object using its extracted pixel shift together with Earth's known orbital velocity (see Section \ref{subsection:orbit_interpretation} for further details of this calculation). We approximated that contributions to the pixel shift from the object's orbital motion were negligible.

\paragraph{Radius}
To calculate estimated radii $r$, we used the scaling relation $F \propto d^{-4}$ for reflected light, where $F$ is flux at the TESS CCD and $d$ is the distance to the object from Earth. Then, the number of counts $N_c$ collected in a pixel is related to the distance $d$ and radius $r$ of an outer solar system object using Equation \ref{eq:NC_proportion}.

\begin{equation}
    N_c \propto \frac{r^2}{d^4}
    \label{eq:NC_proportion}
\end{equation}

Combining this scaling relation with Sedna's known properties and recovered flux, we estimated the radii of all recovered and candidate objects in our pipeline. We used the peak flux obtained for Sedna in its PCA known-path recovery for this scaling, noting that the uncertainty in our radius estimate is directly tied to the uncertainty in an object's true contributed flux. Uncertainties in this contributed flux are introduced by self-subtraction and discrepancies between the object's recovered, linear path and its true path on the sky. By using this scaling, we implicitly adopt the geometric albedo of Sedna, $p_V = 0.32$ \citep{pal2012tnos}.

\paragraph{Flux Calibration}

We scaled all extracted signals with the flux of Sedna in order to deduce the magnitude of each object, as well. This standard scaling relation is given by Equation \ref{eq:Sedna_mag_scaling}. 

\begin{equation}
    \frac{F_1}{F_2} = 10^{(V_2 - V_1)/2.5}
\label{eq:Sedna_mag_scaling}
\end{equation}

We assume that the visual ($V$) magnitude scaling is roughly equivalent to that of the TESS passband, which most closely resembles the $I$ band in the Johnson-Cousins UBVRI system. This assumption would hold in the case of a perfectly spherical, Lambertian surface; however, deviations from sphericity and variations in albedo across objects introduce additional uncertainties to our scaling, and, as a result, to our extracted magnitude estimates. By scaling with Sedna, we again implicitly assume that the albedos of candidate objects are similar to that of Sedna ($p_V = 0.32$). For objects with a true albedo lower than that of Sedna, this means that our predicted sizes may be underestimated.

\begin{figure}
    \centering
    \includegraphics[width=0.45\textwidth]{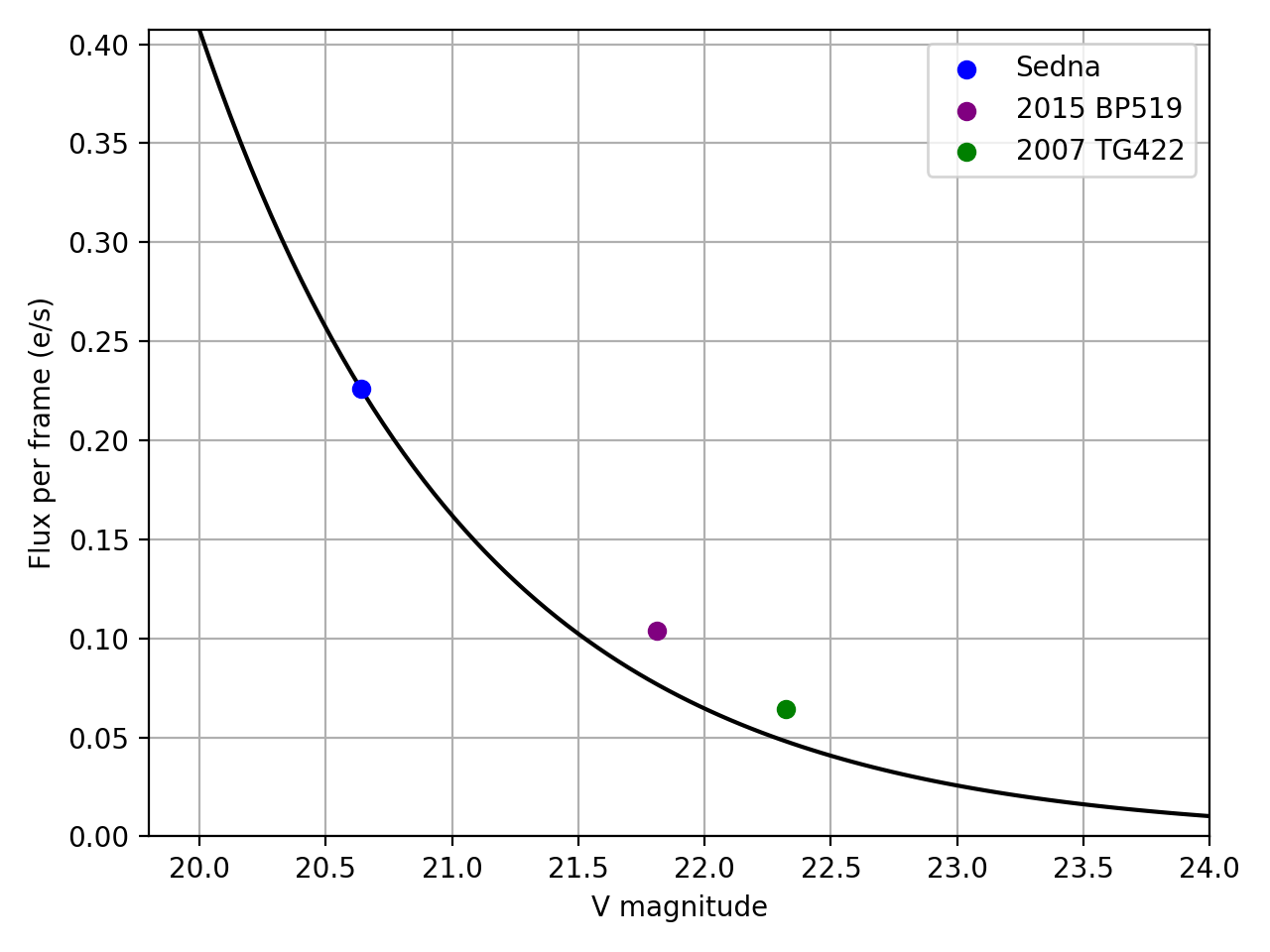}
    \caption{Flux calibration used to estimate the magnitude of all recovered signals through scaling with Sedna. The recovered fluxes of 2015 BP519 and 2007 TG422 are included for reference. The elevated flux per frame of these two objects over the predicted values suggests that true objects may be slightly brighter than our algorithm reports.}
    \label{fig:flux_calibration}
\end{figure}

The results of this scaling are shown in Figure \ref{fig:flux_calibration}, with the known magnitudes and extracted flux values of 2015 BP519 and 2007 TG422 included for reference. We use the known path recoveries to determine the per-frame flux contribution of each object, dividing the peak summed flux by the total number of frames contributing to the sum. This extrapolation leads to an overestimated magnitude for 2015 BP519 and 2007 TG422 given their recovered fluxes, suggesting that real, recovered objects may be slightly brighter than our algorithm reports.

\subsection{Cross-Check Against Known Objects}
After identifying our candidates, we checked all known distant solar system objects to verify whether any should have been observable in our survey. The utility of this exercise is twofold. First, it allows us to determine whether our candidates coincide with previously discovered objects and to verify whether our candidates are truly ``new" potential objects. Second, it provides information regarding whether our algorithm failed to recover any known objects that should have been detectable.

We considered all objects in the \sourcefile{distant_extended.dat} file downloaded from the International Astronomical Union Minor Planet Center\footnote{\url{https://www.minorplanetcenter.net/data}} on February 8th, 2020. This file includes all known TNOs, Centaurs, and scattered disk objects in the Minor Planet Center catalog. We first extracted the sky position of each object as viewed by the TESS spacecraft during the midpoints of Sectors 18 and 19 using the \texttt{jplhorizons} module of \texttt{astroquery} \citep{ginsburg2019astroquery}. Afterwards, we applied the \texttt{TESSCut} module of \texttt{astroquery} to determine whether those coordinates fell within the TESS field of view during Sector 18 and/or 19.

Once we identified a list of objects within the field of view, we then determined whether any were consistent with the limits of our shift-stack search. Our search includes objects moving at a sky-plane projected speed consistent with that of a body on a circular orbit between $d=70-800$ au. More specifically, this means that we searched for objects whose projected speed corresponds to a certain range of pixel shifts: for Sector 18, $4 \leq \Delta x \leq 47$ pixels over a $\sim$19-day temporal baseline, and, for Sector 19, $5 \leq \Delta x \leq 56$ pixels over a $\sim$23-day baseline. In each case, we also allowed for $-8 \leq \Delta y \leq 8$ pixels of movement in the y-direction over the full baseline. The number of pixels traversed by an object at known orbital velocity and distance is described in Section \ref{subsection:orbit_interpretation}. We determined the total $\Delta x$ and $\Delta y$ shifts expected for each object and confirmed that no known objects lie within our search parameter space; instead, they all produce larger $x-$shifts corresponding to smaller geocentric distances than are covered by our search.

\subsection{Time Constraints for Follow-Up}

For the distant solar system objects that we focus on in this work, with short orbital arcs across a single TESS sector, we are sensitive to only the distance and magnitude of recovered objects during the time of the TESS observations. We gain insufficient information from the shift-stacks to place meaningful constraints on the orbital elements of candidates. As a result, candidates must be followed up relatively quickly after the TESS observing sector from which they were extracted.

The sky-plane position of an object on a circular orbit at $d=80$ au, the distance of our most nearby candidate, would change by up to $30\arcmin$ from orbital motion over the span of one year. The most distant candidate at $d=200$ au would shift in position by up to $8\arcmin$. The primary direction of motion can be approximated using the assumption that these objects are on prograde orbits; however, without more detailed orbital information, the candidates would be rapidly lost in the absence of follow-up observations. It may be possible to connect orbital arcs from the current TESS observations with data from the upcoming TESS extended mission, which would allow for a more refined orbit and an extended time available for follow-up observations.

\subsection{Expected Yield}
In total, there are three known distant TNOs -- Sedna, Eris, and Gonggong -- with 70 au $\leq d \leq$ 100 au and $V \lesssim 22$ au. If we assume that these objects are evenly distributed across all TESS sectors, and that any objects in this parameter space lying outside of the galactic plane would have already been detected, we expect that there is a roughly 33\% chance that one of our candidates in that range is real. This approximation comes from the prior that we have searched 2 of 8 sectors crossing through the galactic plane at relatively low inclinations, while roughly 18 sectors lie outside of the galactic plane.

An exact false positive rate is difficult to estimate with this method because our candidate list comprises signals that are not only outliers in flux, but that also appear as point sources in the results from both baseline subtraction methods. While our automated candidate extraction rapidly selects local maxima, it does not distinguish between point sources and more extended flux maxima corresponding to systematics. This final step is conducted through a visual assessment in which the vast majority of flux maxima are dismissed as likely systematics.

Many, if not most, of the high signal significances reported in Table \ref{tab:candidate_list_table} are likely the result of unmodeled systematic errors. Most best-ever frames have several sources detected at $\geq 5\sigma$ significance due to systematic noise. This is because our significance is defined as the deviation above zero attained for a given shift-stack, rather than the deviation above all sources across all possible shift-stacks. In this way, we avoid discarding candidates due to the presence of a strong systematic signal elsewhere in the frame. Despite the large number of flux maxima that are initially recovered, only a small number of these sources pass our visual examination test.

Even with this final vetting step, we anticipate a high false positive rate due to the expected rarity of true objects in this parameter space. For the relatively small sample of objects presented in this work, we propose that the best method to verify the false positive rate would be to follow up the candidates directly with observations.

Future work will extend this search to a more optimized, all-sky survey that incorporates neural networks to rapidly and accurately identify true signals (Rice et al. in prep). By using these neural networks to fully automate the candidate identification process, it will be possible to more rigorously constrain the expected false positive rate for larger-scale surveys.

\newpage
\subsection{Injection Recovery}
\label{subsection:injection_recovery}
We also developed an injection recovery algorithm to study the recoverability of objects with varying magnitudes and path lengths across frames with differing systematics. Each injection consists of a $13\times13$ pixel two-dimensional Gaussian, described by

\begin{equation}
g(x,y) = Ae^{-\Big(\frac{(x-x_0)^2}{2\sigma_x^2} + \frac{(y-y_0)^2}{2\sigma_y^2}\Big)}
\label{eq:gaussian}
\end{equation}
 
We select $\sigma$, which determines the width of the Gaussian, to match the PSF of TESS, where 50\% of light from a point source falls within 1 pix$^2$. We use a symmetric PSF for which $\sigma = \sigma_x = \sigma_y$. The amplitude of the Gaussian injection, $A$, is scaled to match the flux peak expected for an object of a given magnitude based on the extrapolation in Figure \ref{fig:flux_calibration}, multiplied by a factor of 3.5 to account for the fact that the peak recovered flux is shared by roughly 3.5 neighboring pixels. We verified that this scaling provides correctly-calibrated injections by reproducing roughly the expected peak fluxes of the three TNOs described in Section \ref{subsection:recovery_known_objects}.

In our injection tests, we added these signals into the data frames prior to all data pre-processing, then ran our full pipeline with the injected objects included. Afterwards, we applied the same automated source extraction algorithm that we had used in our images with no injections for a direct comparison showing whether these sources, if present in each frame, would have been detected.

\subsubsection{Injection Completeness Tests}

To quantify our recovery rates, we injected grids of 24 identical signals into one $256\times256$ pixel cutout frame from each TESS camera. Signals may be more or less recoverable in different regions of a given best-ever frame due to spatially nonuniform systematics. As a result, we injected many signals into each frame and used the fraction of recovered sources in each frame as an indicator of our recovery rates.

We injected the signal grids into the cutout region with origin (1024, 1024) towards the center of CCD 1 from each camera. Because the on-sky stellar density varies with proximity to the galactic plane, we report results for each camera separately.

\begin{figure*}[h]
    \centering
    \includegraphics[width=0.98\linewidth]{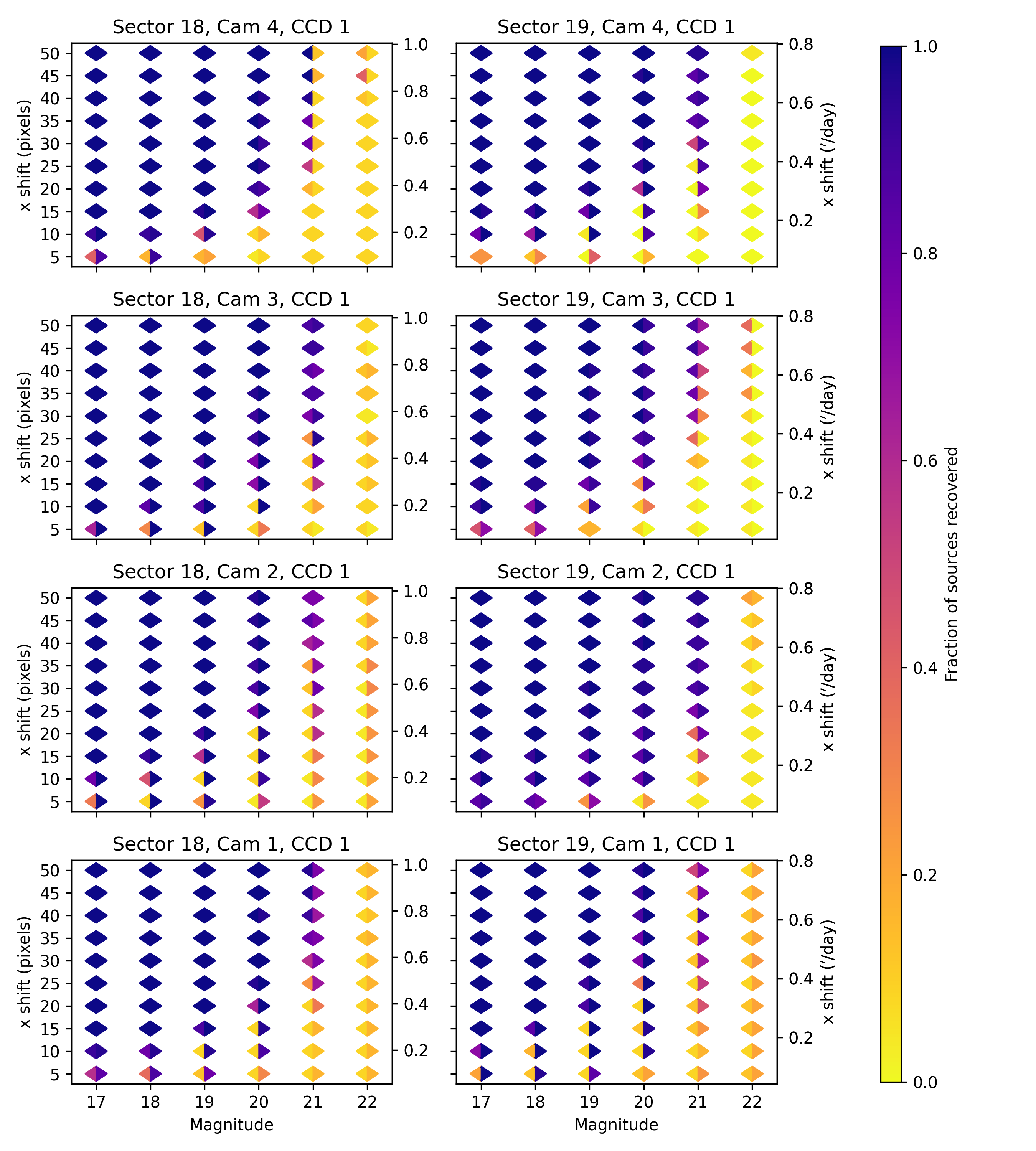}
    \caption{Injection recovery completeness as a function of path length and magnitude for the parameter space explored by our blind search. At each grid point, the left half of the diamond provides the recovery fraction using the polynomial baseline subtraction, while the right half provides the recovery fraction using the PCA baseline subtraction. Each pixel shift directly corresponds to an angular movement provided as an alternative y-axis on the right. For reference, $0.1\arcmin$/day corresponds to an object at $d=550$ au, and $1\arcmin$/day corresponds to $d=50$ au. This angular movement is primarily due to the Earth's parallactic motion, rather than the objects' movement within their own orbits.}
    \label{fig:completeness_grid_5-50}
\end{figure*}

\begin{figure*}
    \centering
    \includegraphics[width=0.98\linewidth]{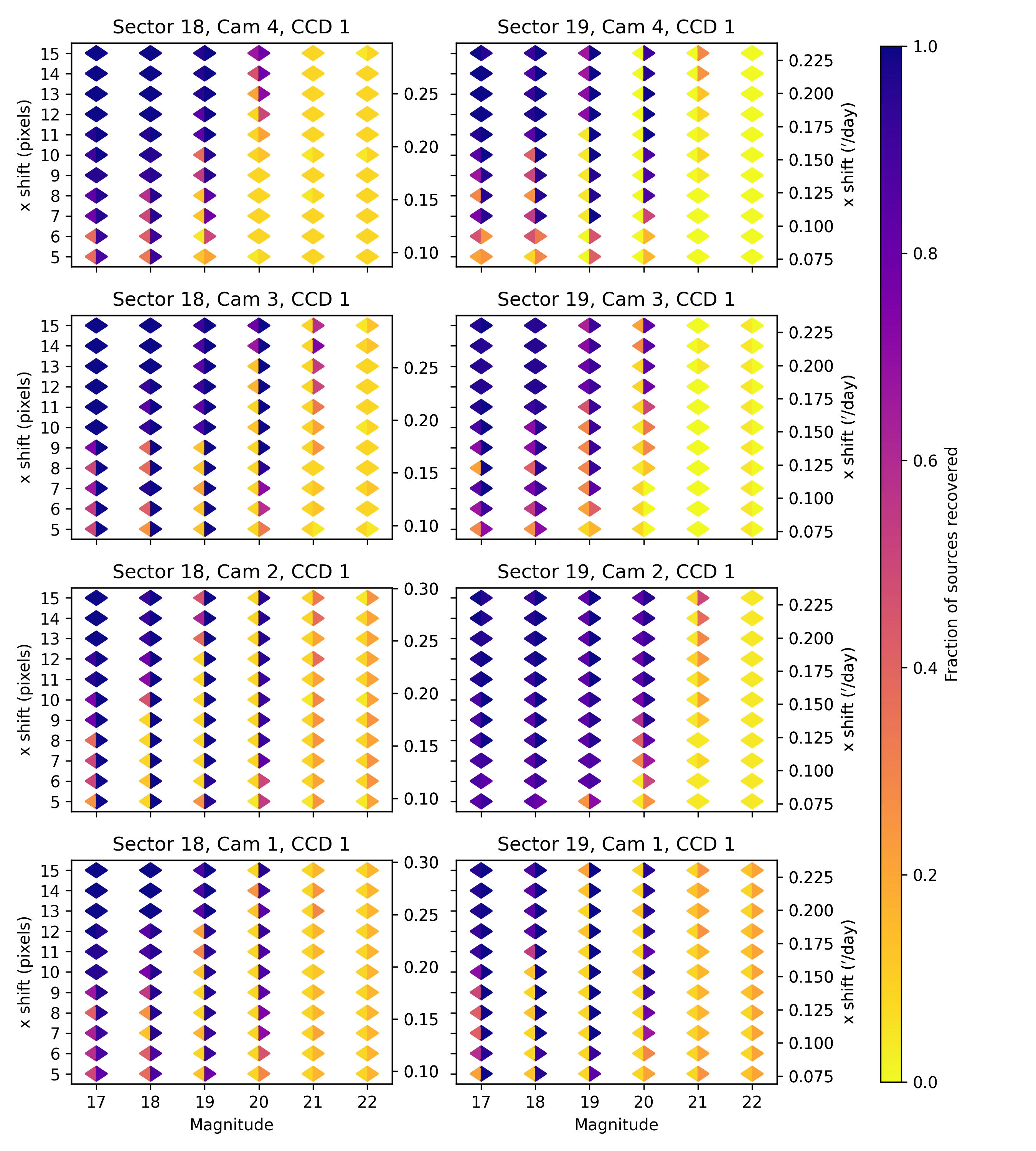}
    \caption{Injection recovery completeness as a function of path length and magnitude for Planet Nine-like orbits. At each grid point, the left half of the diamond provides the recovery fraction using the polynomial baseline subtraction, while the right half provides the recovery fraction using the PCA baseline subtraction. The PCA method is substantially more sensitive to distant, slow-moving objects.}
    \label{fig:completeness_grid_5-15}
\end{figure*}

We varied the magnitudes and path lengths of our injected signals, then determined whether each was recovered by our automated candidate extraction algorithm in the corresponding best-ever frames. For simplicity, all injections in this section move only in the $x-$direction, with zero $y-$shift. We conducted these injection tests across the parameter space of signals explored in this work, testing both the polynomial and PCA recovery methods for comparison. Results from both baseline subtraction methods are provided in Figure \ref{fig:completeness_grid_5-50}. To better understand our survey's sensitivity to Planet Nine, we also conducted injection tests using a finer grid of path lengths spanning the Planet Nine parameter space, with results in Figure \ref{fig:completeness_grid_5-15}.

In a small number of cases, a local maximum had already existed at the location of the injection. In these ambiguous cases, our algorithm accepts the signal as a recovery; this occurs at most 1-2 times in a given camera and therefore cannot artificially inflate our fractional recovery above 10\%. Similarly, lower magnitudes and higher $x-$shifts do not unilaterally lead to higher recovery fractions because, at certain magnitude ranges, the injected signals blend with systematics in the frame and shift the locations of recovered local maxima.

Figure \ref{fig:completeness_grid_5-50} demonstrates that we reliably recover objects with $V<21$ and $x-$shift $>30$ pixels ($d\lesssim150$ au). At shorter path lengths (larger distances) and dimmer fluxes, injected signals are recovered with decreasing consistency. The galactic plane crosses through Camera 1 of Sector 19 and Camera 2 of Sector 18, contributing to a reduced fraction of recoveries in those regions.

The more finely gridded Figure \ref{fig:completeness_grid_5-15} shows that, while our shift-stacking search grid covers much of the parameter space spanned by plausible Planet Nine orbits, the survey presented here cannot rule out the majority of Planet Nine parameter space. This is, in large part, because we require that any accepted candidates are recovered by \textit{both} baseline subtraction methods. Figure \ref{fig:completeness_grid_5-15} demonstrates that the PCA baseline subtraction method consistently produces more reliable recoveries for objects with $d>190$ au -- corresponding to $\Delta x \leq$15 pixels in Sector 18, or $\Delta x \leq$19 pixels in Sector 19 -- and that it is capable of reliably recovering distant ($d\leq800$ au) objects down to $V\sim20$ with a single sector of data alone.

Distant objects remain in the same pixel for a large fraction of the full time series such that they easily blend into the baseline flux. By including only surrounding pixels well outside the central pixel's PSF in the PCA subtraction, we circumvent the polynomial baseline subtraction's tendency to shift the full baseline fit slightly upward in the presence of a very slow-moving object. The PCA method shows great promise to narrow down the parameter space for even objects out to $\sim$800 au ($\Delta x \sim5$ pixels) in future optimizations of this technique.

\subsubsection{Uncertainty Characterization}

We also characterized our uncertainties in the properties of recovered objects using the gridded injection tests. At each magnitude and $x-$shift, we calculated the mean offset of all 24 gridded injections relative to the known values. Then, we determined the standard deviation across the full parameter space displayed in Figure \ref{fig:completeness_grid_5-50}.

We found that the magnitude uncertainties retrieved with the polynomial and PCA baseline subtraction methods were $\sigma_V = 0.7$ and $\sigma_V=1.1$, respectively. The path length uncertainties were characterized by $\sigma_{\Delta x}=13$ pixels and $\sigma_{\Delta y}=2$ pixels for the polynomial subtraction, and $\sigma_{\Delta x}=8$ pixels and $\sigma_{\Delta y}=1$ pixels for the PCA subtraction. These large path length uncertainties, particularly in the $x-$direction of motion, preclude meaningful recoveries of orbital information from our candidates, and they propagate to correspondingly large fractional uncertainties of up to $\sim70\%$ in our estimated candidate distances and radii. The PCA subtraction recovers notably more accurate candidate properties than the polynomial method due to its lack of self-subtraction.

\subsection{Interpretation of Orbits}
\label{subsection:orbit_interpretation} 
Once we have extracted candidates from our best-ever frames, we consider the orbital constraints that can be placed on these objects. In our current framework, these constraints are limited by a few factors -- to keep computation times manageable, our shift-stacking algorithm does not consider sub-pixel shifts, and we assume a straight-line path for the distant solar system objects of interest. For such distant objects, a simplified orbital arc spanning just a few weeks at best does not yield meaningful constraints on the TNOs' true orbits.

Nevertheless, we can set up the formalism to understand what types of projected tracks correspond to plausible orbits and which correspond instead to either physically impossible or highly implausible tracks. The formalism described here demonstrates how a set of orbital elements can be translated to $x-$ and $y-$ TESS pixel shifts by determining the angular sky-plane path associated with the orbit. Our translation to pixel shifts is TESS-specific; however, we note that \citet{bernstein2000orbit} has also outlined a general framework to approximate orbital properties in the short-arc limit, and we refer the interested reader to this study for details.

We use standard coordinate transformations to translate the orbital elements into Cartesian space. We complete two transformations: first, a transformation to the heliocentric ecliptic reference frame, and then to a barycentric frame that aligns with the TESS FFIs for ease of interpretation. The first of these transformations is given by Equation \ref{eq:transform_heliocentricecliptic}, where we deviate from the standard labels for each direction in order to maintain consistency with the 2D $xy$ directions that we use to describe locations in the TESS FFIs. Throughout this section, we use the capitalized $XYZ$ notation to describe Cartesian distances, while we continue to refer to TESS pixel shifts using the lower-case $xy$ notation.

\begin{equation}
\begin{pmatrix} Z_h \\ X_h \\ Y_h \\ \end{pmatrix} =
r \begin{pmatrix}
\cos\Omega\cos(\omega+f) - \sin\Omega\sin(\omega+f)\cos i \\
\sin\Omega\cos(\omega+f) + \cos\Omega\sin(\omega+f)\cos i\\
\sin(\omega + f)\sin i
\end{pmatrix}
\label{eq:transform_heliocentricecliptic}
\end{equation}

In this heliocentric ecliptic reference frame, $Z_h$ is defined to be perpendicular to the ecliptic plane, while $X_h$ points in the direction of the vernal equinox and $Y_h$ is mutually perpendicular. We then switch to a custom geocentric frame with Equation \ref{eq:transform_customgeocentric} using a translational transformation to Earth's location, as well as one final rotational transformation about the $Y_h$ axis using the angle $\phi$ between the vernal equinox and the Sun-Earth vector at the central time of a sector.

\begin{equation}
\begin{pmatrix} Z \\ X \\ Y \\ \end{pmatrix} =
\begin{pmatrix} 
Z_h \cos\phi - X_h \sin\phi \\
Z_h \sin\phi + X_h \cos\phi \\
Y_h \end{pmatrix} + 
\begin{pmatrix}
    d_{ES}\cos\phi \\ d_{ES}\sin\phi \\ 0
\end{pmatrix}
    \label{eq:transform_customgeocentric}
\end{equation}

Here, $d_{ES}$ is the length of the Sun-Earth vector. While this coordinate system is centered on the Earth rather than the TESS spacecraft itself, the difference between the two is negligible for our purposes. In this final coordinate frame, $Z$ points in the radial direction perpendicular to the TESS FFI frames. The $X$ and $Y$ directions are aligned with the TESS CCDs, where $X$ is parallel to the ecliptic plane in the projected frames (consistent with our notation for the $x$-pixel shift direction), while $Y$ is perpendicular to the ecliptic plane (in the $y$-shift direction).

Using these steps, we can obtain the Cartesian distances traveled by each object in 3D space along a specific orbital arc. Then, we can translate those distances into a projected sky-plane track as observed from Earth. For a known orbit, this tells us exactly where an object is in the TESS frame at each time. 

The $y$-component of a TNO's track across the detector should include no contribution from Earth's parallax, since the TESS frames are aligned with the ecliptic plane (by definition, the plane of Earth's orbit). Thus, any observed $y$-shift of a TNO in the TESS CCDs is due to orbital motion. Large $y$-shifts must correspond to objects on orbits with high inclination and with locations on their orbits that bring them relatively close to Earth -- either through a high-eccentricity orbit where the detected object is near periastron, or through a small semimajor axis.

The projected angular movement of an object in the $y$-direction across TESS's CCDs is provided by $\theta_Y$, which is described by 

\begin{equation}
    \theta_Y = \frac{\Delta Y}{Z}\, .
\label{eq:theta_Y}
\end{equation}

Here, $Z$ is the mean line-of-sight distance between the start and end of the orbital arc. This radial distance should be much smaller than the radial movement $\Delta Z$ of the object across the TESS single-sector time baseline ($Z << \Delta Z$; $\Delta t = t_2 - t_1 \sim 27$ days for one sector). The movement of the object over $\Delta t$ in the $Y-$direction, defined as perpendicular to the ecliptic plane, is given by $\Delta Y$. We assume that $\Delta Y < < Z$ such that $\tan\theta_Y \sim \sin\theta_Y \sim \theta_Y$. The resulting $\theta_Y$ from Equation \ref{eq:theta_Y} provides the object's vertical projected movement in units of radians. 

Given that TESS's pixels span $21\arcsec\times21\arcsec$, we directly convert these angular sky movements to projected pixel shifts on the TESS detector using Equation \ref{eq:N_pix_shift}.

\begin{equation}
    N = \theta \hspace{1mm} \times \frac{206265\arcsec}{1 \hspace{1mm} \mathrm{rad}} \times \frac{1 \hspace{1mm} \mathrm{pix}}{21\arcsec}
\label{eq:N_pix_shift}
\end{equation}

The projected horizontal angular movement of an object across the detector, $\theta_X$, can also be extracted from the object's known orbit. In this direction, parallax contributes directly to the path length spanned by an object. Earth moves in its orbit at $v_{\oplus} = 29.78$ km/s, and its velocity in the ecliptic plane relative to that of a candidate object $v_c$ contributes to the candidate's $\theta_X$ path length. For simplicity, we approximate that the velocity vectors of Earth and the observed object are roughly parallel at the time of observation, meaning that the net horizontal angular shift $\theta_X$ from both parallax and orbital motion is 

\begin{equation}
    \theta_X = \frac{\Delta X + (v_{\oplus} - v_c)\Delta t}{Z}.
\label{eq:theta_X}
\end{equation}

Then, we can again apply Equation \ref{eq:N_pix_shift} to convert this angular shift to a pixel shift in the TESS frames, providing the $x$- pixel shift for a given orbital arc. 

This translation between orbital elements and pixel shifts will be useful to forward model orbit bundles consistent with a given orbital track. While we do not have strong enough constraints on the outer solar system objects in this study to warrant such an analysis, this type of forward modeling will be applicable to extensions of this survey tracking shorter-period orbits, where a larger fraction of the orbit is spanned by the TESS observations. In the shorter-period regime, shift-stacking may serve as a useful tool to refine the orbits of known objects.

\section{Discussion}
\label{section:discussion}

\subsection{Expected Yield of an All-Sky Survey}
There is a marked difference between the systematics present in, for example, Camera 1 of Sector 19 -- located directly along the galactic plane -- and Camera 4 of Sector 19, at the northern ecliptic pole with $\gtrsim30\degr$ separation from the galactic plane. As expected, we find that our algorithm returns a much cleaner baseline subtraction for frames far from the galactic plane than those pointed towards the plane and neighboring regions. This is due to stellar crowding in the vicinity of the galactic plane, which has stymied previous optical searches in this region. Roughly 30\% of all TESS frames lie either directly along or adjacent to the galactic plane.

The three objects presented in \citet{holman2019tess} -- Sedna, 2015 BP519, and 2015 BM518 -- each also have $\gtrsim30\degr$ separation from the galactic plane. This suggests that the yield estimates in \citet{payne2019tess} may be optimistic, since they are based upon these three objects that are located in particularly ``clean" regions of the sky. However, our algorithm recovers both Sedna and 2015 BP519 at significantly higher SNR than the recoveries presented in \citet{holman2019tess}, suggesting that the magnitude limits in these ``clean" regions may be even lower than previously estimated. An all-sky shift-stacking search using the TESS FFIs should have varying sensitivity as a function of proximity to the galactic plane.

We stress that even within Sectors 18 and 19, our search is not exhaustive. As demonstrated by our detection recovery grids in Section \ref{subsection:injection_recovery}, the two baseline subtraction methods applied throughout this work are not equally sensitive to distant TNOs. We ultimately find that the PCA baseline subtraction method, because of its robustness against self-subtraction, is more capable of narrowing down the Planet Nine parameter space and discovering new distant TNOs than the polynomial baseline subtraction method. While we required in this work that our candidates were recovered by both methods, this may not be necessary in future work. Extensions of this project may instead consider using only one baseline subtraction method, incorporating data from the ongoing TESS extended mission that is in progress at the time of publication for an additional check.

Hardware limitations also place some minor limits on the completeness of our search. A gap is left between each TESS camera, and the four CCDs in each camera are separated by 2 mm each. As described in Section \ref{subsection:new_candidates}, some frames in CCD 3 of Camera 3 also show substantial systematic effects due to saturation from the bright star Polaris, dramatically reducing our sensitivity in the small subset of our frames containing that column. However, while these regions of negligible sensitivity exist within our search, the likelihood is slim that a rare object within our magnitude limit falls within these regions.

\subsection{Future Directions}
The analysis presented here focuses on only two TESS sectors -- a small fraction of the full TESS dataset's vast sky coverage. This work can be naturally extended to a larger-scale survey by incorporating convolutional neural networks (CNNs) into the existing pipeline to streamline the process of extracting promising candidates from the pre-processed best-ever images. These CNNs can be trained on injected Gaussian signals corresponding to dim objects at a known magnitude and distance, which will simultaneously provide a rigorous test of our algorithm's performance in various regions of the sky. Automating the candidate vetting process allows for a similar search on a much larger scale, incorporating the full TESS FFI dataset to survey almost the entire sky. Future work will explore this in greater detail (Rice et al., in prep).

Beyond its originally planned 2-year survey, TESS has been approved for an extended mission that will again survey nearly the full sky, but at a 10-minute cadence rather than a 30-minute cadence. When combined with the original survey, the longer temporal baseline provided by the extended mission can be leveraged to more efficiently search for slow-moving objects in the very distant solar system. While enabling higher signal-to-noise detections and accordingly lower magnitude limits for all TNOs, this additional data will be especially useful in searches for objects as distant as the predicted Planet Nine. 

Objects located hundreds of au from the Earth are particularly prone to self-subtraction, since they remain in an individual pixel for a large fraction of the full time series and therefore easily blend into the baseline flux. The longer temporal baseline afforded by the TESS extended mission will help to push the detection limits of very distant TNOs by increasing their total path lengths in the TESS dataset. Furthermore, the two independent sets of observations may be analyzed separately, and the combined results can be used to reject false positive signals. This would make it possible to use only the PCA baseline subtraction method, which is sensitive to a wider range of objects than the polynomial method, throughout the analysis rather than requiring two separate methods.

\subsection{Additional Applications}

\subsubsection{Small Bodies Interior to 70 au}
While this work focuses specifically on the distant solar system, the same algorithm can also be applied to study solar system objects interior to 70 au. \citet{payne2019tess} identifies Centaurs, as well as high-inclination Kuiper belt objects (including those interior to 70 au) as some of the most promising objects to search for with TESS shift-stacking. Other high-inclination objects interior to the Kuiper belt would be possible to study, as well; however, for more nearby objects, orbital motion comprises a larger component of the targets' total sky-plane motion. This means that the approximation that Earth's parallax dominates the objects' motion no longer holds. As a result, surveys looking for objects interior to our current search limit will require additional planning to search along specified orbits, rather than along straight paths across the detector. 

Examining the full range of possible nonlinear orbits is more computationally expensive than a linear path search. Nevertheless, the problem is still tractable. \citet{burkhart2016deep} found that, using the methods outlined in \citet{parker2010pencilnovel}, only $\sim$35 total paths were required to conduct an exhaustive search for satellites of Haumea lying 10,000-350,000 km from the primary. This is because only sufficiently distinct sky tracks need to be searched in order to recover sources along all possible orbits within a set error tolerance. Beyond this threshold, similar tracks with substantial overlap provide diminishing returns.

\subsubsection{Interstellar Objects}
Given that the TESS field of view probes the high-inclination solar system, shift-stacking with TESS FFIs may be a powerful tool to search for interstellar objects (ISOs) and place limits on their occurrence rate. The two ISOs confirmed so far -- 1I/'Oumuamua and 2I/Borisov -- peaked in flux at $V\sim 20$ and $V\sim 15$, respectively, both comfortably within our magnitude limits for a single-sector shift-stack search \citep{meech2017, guzik2020initial}. Using the TESS continuous viewing zones at the ecliptic poles, these magnitude limits could be pushed even lower. With 13 sectors of data included in a single analysis, our magnitude limit would be extended to $V\sim24-25$ for an object remaining in the field for the full duration of these observations. The discovery of both 1I/'Oumuamua and 2I/Borisov in relatively rapid succession suggests an abundance of ISOs passing through the solar neighborhood \citep{rice2019hidden}, and a deep study with TESS could place stringent constraints on the occurrence rate and size distribution of these rare objects. 

Each of our pipeline components has been designed with flexibility and interchangeability in mind, meaning that our algorithm can be adapted for application to different datasets with relative ease. The \textit{Kepler} dataset \citep{borucki2010kepler} may be a particularly powerful probe of ISO occurrence rates given its extended temporal baseline and its pointing towards the galactic apex -- the direction from which interstellar objects are most likely to approach the solar system. Although the initial treatment of systematics would differ for the \textit{Kepler} dataset, the remainder of our pipeline could be easily applied to this new set of images given an adjusted, \textit{Kepler}-specific baseline subtraction module.

Beyond a statistical study of ISOs, a shift-stacking survey with TESS FFIs may also be capable of finding individual ISOs for further follow-up. While TESS data is not publicly released until months after observing, it may be possible to discover individual objects that have been observed by TESS on their way towards perihelion. Because individual ISOs are typically observable only for a short span of time, this survey would require a rapid turnaround time for follow-up observations once candidates have been identified. This may be possible with a fully automated pipeline that incorporates CNNs.

\subsubsection{Directly Imaged Planets}
\citet{males2013direct} have demonstrated that orbital motion will be significant over the integration times needed to directly image habitable-zone planets in extrasolar systems, suggesting that ``de-orbiting" prospective planets over many trial orbits, while leading to an increase in false alarms, will also be necessary for survey completeness. By shift-stacking along theoretical orbits, it is possible to recover an object's signal at its nominal value without leakage from orbital motion. Thus, particularly in the era of large telescopes, shift-stacking may also provide an important tool to fully utilize the information content of direct imaging surveys.

\section{Conclusions}
\label{section:conclusion}
We have developed a novel pipeline that is custom-designed to search for outer solar system objects by shift-stacking FFIs from the TESS dataset. In this paper, we highlighted the performance of this pipeline by recovering three known TNOs down to $V\sim22$. We then applied the pipeline to two sectors of TESS data -- Sectors 18 and 19 -- located along the northern galactic plane in a targeted search for Planet Nine and other extreme trans-Neptunian objects. From this search, we extracted a list of promising candidates that can be easily checked with optical follow-up observations using meter-class telescopes.

This work serves as a proof-of-concept that develops the foundation for larger-scale applications of a similar technique. The existing framework applied in this work can reliably blindly recover the signals of distant solar system bodies in the galactic plane with V$<21$ and current distances $d\lesssim 150$ au. These limits are currently set by the methodology applied -- that is, the requirement that candidates are recovered using two separate methods, one of which is less sensitive than the other -- rather than the dataset itself, indicating that the physical boundaries of our detection limits have not yet been met. The sensitivity of our survey also improves with distance from the galactic plane. Future optimizations of this framework, including the incorporation of neural networks and of additional data from the TESS extended mission, will further push the boundaries of these detection limits and the range of solar system bodies that can be discovered and studied with TESS.

\section{Acknowledgements}
\label{section:acknowledgements}

We thank Matt Payne and Darin Ragozzine for discussions that have helped to refine the ideas explored in this work. We also thank the anonymous referees for thoughtful comments and suggestions that substantially improved this manuscript. M.R. is supported by the National Science Foundation Graduate Research Fellowship Program under Grant Number DGE-1752134. This material is based upon work supported by the National Aeronautics and Space Administration through the NASA Astrobiology Institute under Cooperative Agreement Notice NNH13ZDA017C issued through the Science Mission Directorate. We acknowledge support from the NASA Astrobiology Institute through a cooperative agreement between NASA Ames Research Center and Yale University. This research has made use of data and/or services provided by the International Astronomical Union's Minor Planet Center. We thank the Yale Center for Research Computing for use of the research computing infrastructure. This project was developed in part at the Expanding the Science of TESS meeting, which took place in 2020 February at the University of Sydney.

\software{\texttt{numpy} \citep{oliphant2006guide, walt2011numpy, harris2020array}, \texttt{matplotlib} \citep{hunter2007matplotlib}, \texttt{lightkurve} \citep{lightkurve2018}, \texttt{astroquery} \citep{ginsburg2019astroquery}, \texttt{PyEphem} \citep{rhodes2011pyephem}}, \texttt{astropy} \citep{astropy2013, astropy2018}, \texttt{scipy} \citep{virtanen2020scipy}

\bibliography{bibliography}
\bibliographystyle{aasjournal}

\end{document}